\documentclass[11pt, a4paper]{article}
\usepackage{rotating}
\usepackage{authblk}

\usepackage[top=2cm, bottom=2cm, left=2cm, right=2cm, footnotesep=1cm]{geometry}

\usepackage[authoryear,round]{natbib}
\usepackage{graphicx}
\usepackage{subcaption}
\usepackage{amsthm}
\usepackage[justification=centering]{caption}
\usepackage{tabulary}
\usepackage{setspace}
\usepackage{amsmath}
\usepackage{siunitx}
\usepackage{multirow}
\usepackage{mleftright}
\usepackage[nocomma]{optidef}
\usepackage[titletoc,title]{appendix}
\usepackage[linesnumbered,ruled,vlined]{algorithm2e}
\usepackage{algpseudocode}
\usepackage{pdflscape}
\usepackage{booktabs}
\usepackage{array}
\usepackage{tabularx}
\usepackage{adjustbox}
\usepackage{afterpage}
\usepackage{placeins}
\usepackage{float}
\usepackage{xcolor}

\usepackage[colorlinks,linkcolor=black,anchorcolor=black,citecolor=black,urlcolor=black]{hyperref}
\usepackage{lineno}
\usepackage[shortlabels]{enumitem}
\usepackage{palatino}
\usepackage{newpxtext}
\usepackage{newpxmath}
\usepackage{soul}
\usepackage{longtable}
\usepackage{stackengine}
\usepackage{titlesec}
\usepackage{etoolbox}
\titlespacing*{\section}{0pt}{1.75ex plus .5ex minus .1ex}{1.15ex plus .1ex}
\titlespacing*{\subsection}{0pt}{.75ex plus .1ex}{.75ex plus .1ex}
\titlespacing*{\subsubsection}{0pt}{1.625ex plus .5ex minus .1ex}{.75ex plus .1ex}
\makeatletter
\patchcmd{\@maketitle}{\vskip 2em}{}{}{}
\makeatother
\makeatletter
\AtBeginDocument{%
  \g@addto@macro\normalsize{%
    \setlength\abovedisplayskip{5.5pt plus 1.5pt minus 3pt}%
    \setlength\belowdisplayskip{5.5pt plus 1.5pt minus 3pt}%
    \setlength\abovedisplayshortskip{0pt plus 1.5pt}%
    \setlength\belowdisplayshortskip{3.25pt plus 1.75pt minus 1.5pt}%
  }%
  \normalsize
}
\makeatother

\graphicspath{{figs_in_paper/}}

\title{Group boarding for airplanes: benchmarking static policies and optimizing dynamic assignment with deep reinforcement learning}

\author[a]{Minyu Shen}
\author[b]{Weihua Gu}
\author[c]{Junqi Ma}
\author[a]{Boqian Song}
\author[b]{Li Zhen\footnote{Corresponding author. Email: l-i.zhen@connect.polyu.hk}}
\author[c]{Gang Kou}

\affil[a]{School of Management Science and Engineering, Southwestern University of Finance and Economics, China}
\affil[b]{Department of Electrical and Electronic Engineering, The Hong Kong Polytechnic University}
\affil[c]{School of Business Administration, Southwestern University of Finance and Economics, China}
\date{\vspace{-4em}}

\begin{document}
\maketitle
\noindent\hrulefill

\onehalfspacing
\captionsetup[figure]{font=normalsize, aboveskip=0pt, belowskip=0pt}
\captionsetup[table]{font=normalsize}
\setlength{\textfloatsep}{0.5\baselineskip}
\setlength{\floatsep}{0.5\baselineskip}
\setlength{\intextsep}{0.5\baselineskip}

\section*{Abstract}

Improving boarding efficiency reduces airplane turnaround time and improves passenger experience. Airlines typically assign passengers to a few sequential boarding groups using static seat-based rules. Yet arrivals, seat choices, and luggage are sequential and random, and a static rule ignores the seats earlier passengers have already taken. We propose the first dynamic formulation of boarding group assignment. As each passenger checks in, we observe earlier passengers' seats and groups, the current passenger's seat, and optional luggage information, then assign a group while keeping companions together. We formulate dynamic group assignment as a Markov decision process and solve it with reinforcement learning (RL). The policy uses a convolutional neural network to encode the checked-in seat-assignment state and is trained by proximal policy optimization. The reward balances total boarding time and average individual boarding time.

We benchmark the proposed RL policy against three companion-compatible static policies (back-to-front, modified Steffen, and alternating block) in an in-house simulator covering six single- and double-aisle layouts. Back-to-front with optimized group sizes achieves the shortest total boarding time and average individual boarding time among the static benchmarks across all layouts. The dynamic RL policy further outperforms it on both metrics in every layout. On a representative case, the RL policy outperforms the optimal back-to-front by up to 9.8\% in total boarding time and 22.8\% in average individual time. Sweeping the reward weight yields an approximate Pareto frontier for operator choice. Trained policies remain robust under out-of-distribution operating conditions, including varying load factors, companion sizes, and luggage loads.

\vspace{0.75em}
\noindent \textbf{Keywords:} Airplane boarding; Boarding group; Travel companion; Group assignment; Deep reinforcement learning.

\section{Introduction}

Turnaround time, the time an airplane spends on the ground between two consecutive flights, includes cleaning, refueling, maintenance, baggage handling, and passenger boarding \citep{jaehn2015airplane}. A Boeing study found that for flights averaging 500 nautical miles, cutting turnaround time by 10 minutes raises airplane utilization by 8 percent \citep{boeing2011impact}. Among these activities, passenger boarding is a major bottleneck \citep{neumann2019boarding}, and unlike operations such as refueling, it offers substantial room for improvement.\par

Boarding matters for two reasons, captured by two metrics. The \textit{total boarding time}, from the first passenger's entry until the last passenger is seated, measures operational efficiency and feeds directly into turnaround time. The \textit{average individual boarding time}, the mean time from the start of boarding until a passenger is seated, measures the passenger experience. The two can conflict: a boarding policy that shortens one may lengthen the other \citep{bachmat2023air}. Boarding policies should therefore be judged on both, not on total boarding time alone, as most prior work has done.\par

Two types of interference impede the boarding process: \textit{aisle interference} and \textit{seat interference} \citep{bazargan2007linear,jaehn2015airplane,willamowski2022minimizing,erland2024let}. Aisle interference occurs when a passenger blocks the aisle while stowing luggage in the narrow cabin, so passengers behind cannot reach their rows. Seat interference occurs when seated passengers must leave their seats to let another passenger reach an inner seat, for example when a window-seat passenger passes others already seated in the aisle and middle seats. The passengers who stand up wait in the aisle until the new passenger sits down, which blocks the row and the aisle and can delay passengers behind.\par

Airlines reduce these interferences by sorting passengers into boarding groups that board in sequence. The group should be decided at check-in and printed on the boarding pass. It can be determined using the information received by the airline, including the passenger's chosen seat, carry-on luggage declaration, and whether the passenger travels with companions. Check-in is also sequential: passengers arrive one at a time in no fixed order, and their assigned group must be set on arrival, before information about later passengers is known.\par

In practice, though, the boarding group is set by a rule fixed in advance, the same no matter who checks in or when. The companion-compatible static benchmarks considered in this paper are adapted from back-to-front \citep{bachmat2013optimal}, modified-Steffen \citep{steffen2008optimal}, and alternating-block \citep{van2002reducing}; they group passengers by row and, at most, by cabin side. This design keeps adjacent companions together, but it leaves unused the seat column, which sets seat interference, and the luggage, which sets how long the aisle stays blocked \citep{coppens2018improving,erland2024let}. The methods that do exploit the column, such as outside-in and reverse-pyramid, cut seat interference sharply but only by separating companions. Optimization-based work goes beyond fixed row rules by exploiting richer seat and passenger information, including row, column, carry-on luggage, and, in some cases, companion constraints \citep{van2005america,bazargan2007linear,soolaki2012new,milne2016optimization,salari2019airplane}, yet it solves an offline boarding or seating problem after the relevant passenger set is known, too late to tell each passenger at check-in. The closest online-adaptive work is \citet{notomista2016fast}, which assigns seats online based on passenger classification; in contrast, our setting keeps seat assignments fixed and determines boarding groups dynamically at check-in. No method, then, combines all three: optimizing state-dependent group assignment at check-in, using seat-column and luggage information, and keeping companion groups intact (see Table~\ref{table:literature_review}).\par

\begin{landscape}
{\small
\begin{longtable}{
>{\raggedright\arraybackslash}p{3.3cm}
>{\raggedright\arraybackslash}p{1.8cm}
*{2}{>{\centering\arraybackslash}p{1.4cm}}
*{2}{>{\centering\arraybackslash}p{0.5cm}}
*{3}{>{\centering\arraybackslash}p{1.45cm}}
>{\raggedright\arraybackslash}p{4.2cm}}

\caption{Summary of the Literature Review}\label{table:literature_review} \\
\toprule
Authors (year) & Methodology & \multicolumn{2}{c}{Airplane layout\textsuperscript{a}} & \multicolumn{2}{c}{Interference} & \multicolumn{3}{c}{\shortstack{Additional assignment\\considerations\textsuperscript{b}}} & Policies evaluated\\
\cmidrule(lr){3-4} \cmidrule(lr){5-6} \cmidrule(lr){7-9}
 & & Single-aisle & Double-aisle & Aisle & Seat & \shortstack{Com-\\panions} & Luggage & \shortstack{Check-in-time\\group\\assignment} & \\
\midrule
\endfirsthead
\multicolumn{10}{c}{\tablename~\thetable{} -- continued from previous page}\\
\toprule
Authors (year) & Methodology & \multicolumn{2}{c}{Airplane layout\textsuperscript{a}} & \multicolumn{2}{c}{Interference} & \multicolumn{3}{c}{\shortstack{Additional assignment\\considerations\textsuperscript{b}}} & Policies evaluated\\
\cmidrule(lr){3-4} \cmidrule(lr){5-6} \cmidrule(lr){7-9}
 & & Single-aisle & Double-aisle & Aisle & Seat & \shortstack{Com-\\panions} & Luggage & \shortstack{Check-in-time\\group\\assignment} & \\
\midrule
\endhead
\citet{van2002reducing} & Simulation & (3+3)*23 &  & \checkmark & \checkmark &  &  &  & BF, AB, OI, OTHERS, PRECISE \\
\addlinespace
\citet{van2005america} & Optimization & (3+3)*23 &  & \checkmark & \checkmark & \checkmark &  &  & BF, RP, group policy by MINLP \\
\addlinespace
\citet{ferrari2005robustness} & Simulation & (3+3)*23 &  & \checkmark & \checkmark &  &  &  & BF, AB, OI, SG, RP, OTHERS, PRECISE \\
\addlinespace
\citet{bachmat2005airplane} & Physics model & \multicolumn{2}{c}{generic} & \checkmark &  &  &  &  & BF, OI, SG \\
\addlinespace
\citet{bachmat2006analysis} & Physics model & \multicolumn{2}{c}{generic} & \checkmark &  &  &  &  & BF, OI \\
\addlinespace
\citet{bazargan2007linear} & Optimization & (3+3)*23 &  & \checkmark & \checkmark & \checkmark &  &  & BF, OI, RP, group policy by MILP \\
\addlinespace
\citet{steffen2008optimal} & Simulation & (3+3)*20 &  & \checkmark &  &  &  &  & BF, FB, Steffen, OI, PRECISE \\
\addlinespace
\citet{schultz2008efficiency} & Simulation & (3+3)*29 &  & \checkmark & \checkmark &  &  &  & BF, AB, OI \\
\addlinespace
\citet{bachmat2009analysis} & Physics model & \multicolumn{2}{c}{generic} & \checkmark &  &  &  &  & BF, AB, OI, OTHERS \\
\addlinespace
\citet{steiner2009speeding} & Simulation & (3+3)*33 &  & \checkmark & \checkmark &  &  &  & BF \\
\addlinespace
\citet{audenaert2009multi} & Simulation & (3+3)*23 &  & \checkmark & \checkmark &  & \checkmark &  & BF, AB, OI, RP, OTHERS, PRECISE \\
\addlinespace
\citet{steffen2012experimental} & Field experiment & (3+3)*12 &  & \checkmark & \checkmark &  &  &  & BF, AB, Steffen, OI, PRECISE \\
\addlinespace
\citet{tang2012aircraft} & Simulation & (3+3)*25 &  & \checkmark & \checkmark &  & \checkmark &  & OTHERS, PRECISE \\
\addlinespace
\citet{soolaki2012new} & Optimization & (3+3)*23 &  & \checkmark & \checkmark &  &  &  & RP, group policy by MILP \\
\addlinespace
\citet{bachmat2013optimal} & Physics model & \multicolumn{2}{c}{generic} & \checkmark &  &  &  &  & BF \\
\addlinespace
\citet{budesca2014optimization} & Simulation & (3+3)*30 &  & \checkmark & \checkmark & \checkmark &  &  & BF, FB, OTHERS \\
\addlinespace
\citet{milne2014new} & Simulation & (3+3)*20 &  & \checkmark & \checkmark &  & \checkmark &  & BF, Steffen, OTHERS, PRECISE \\
\addlinespace
\citet{milne2016optimization} & Optimization & (3+3)*20 &  & \checkmark &  &  & \checkmark &  & Steffen, PRECISE \\
\addlinespace
\citet{notomista2016fast} & Simulation & (3+3)*30 &  & \checkmark & \checkmark & \checkmark & \checkmark &  & OTHERS \\
\addlinespace
\citet{giitsidis2016modeling} & Simulation & (3+3)*23 & {\scriptsize\shortstack{12 seats/row\\$\ast$48}} & \checkmark & \checkmark &  &  &  & BF, AB, OI, SG, RP \\
\addlinespace
\citet{zeineddine2017dynamically} & Simulation & (3+3)*20 &  & \checkmark & \checkmark & \checkmark &  &  & BF, Steffen, OI, OTHERS, PRECISE \\
\addlinespace
\citet{jafer2017comparative} & Simulation & (3+3)*23 &  & \checkmark & \checkmark &  &  &  & BF, AB, Steffen, OI, RP, OTHERS, PRECISE \\
\addlinespace
\citet{ren2018experimental} & Field experiment & (3+3)*8 &  & \checkmark & \checkmark &  &  &  & BF, OI, RP, OTHERS \\
\addlinespace
\citet{delcea2018investigating} & Simulation & (3+3)*29 &  & \checkmark & \checkmark &  &  &  & BF, OI, RP, OTHERS, PRECISE \\
\addlinespace
\citet{milne2018robust} & Optimization & (3+3)*20 &  & \checkmark &  &  & \checkmark &  & Steffen, PRECISE \\
\addlinespace
\citet{tang2018aircraft} & Simulation & (3+3)*25 &  & \checkmark & \checkmark &  &  &  & OTHERS \\
\addlinespace
\citet{schultz2018implementation} & Simulation & (3+3)*29 &  & \checkmark & \checkmark &  & \checkmark &  & BF, AB, OI, RP, PRECISE \\
\addlinespace
\citet{schultz2018advancements} & Simulation & (3+3)*29 &  & \checkmark & \checkmark & \checkmark & \checkmark &  & BF, OI \\
\addlinespace
\citet{schultz2018field} & Field experiment & (3+3)*29 &  & \checkmark & \checkmark &  &  &  & BF, OI \\
\addlinespace
\citet{bachmat2019airplane} & Physics model & \multicolumn{2}{c}{generic} & \checkmark &  &  & \checkmark &  & slow-/fast-first, OTHERS \\
\addlinespace
\citet{erland2019lorentzian} & Physics model & \multicolumn{2}{c}{generic} & \checkmark &  &  & \checkmark &  & slow-/fast-first \\
\addlinespace
\citet{cotfas2019testing} & Simulation & (3+3)*30 &  & \checkmark & \checkmark &  &  &  & BF, RP, OTHERS \\
\addlinespace
\citet{milne2019new} & Simulation & (3+3)*30 &  & \checkmark & \checkmark &  &  &  & BF, RP, OTHERS \\
\addlinespace
\citet{salari2019airplane} & Optimization & (3+3)*20 &  & \checkmark & \checkmark &  & \checkmark &  & Steffen, PRECISE \\
\addlinespace
\citet{kisiel2020resilience} & Simulation & (3+3)*30 &  & \checkmark & \checkmark &  &  &  & BF, Steffen, OI, OTHERS \\
\addlinespace
\citet{erland2021lorentzian} & Physics model & \multicolumn{2}{c}{generic} & \checkmark &  &  & \checkmark &  & slow-/fast-first \\
\addlinespace
\citet{willamowski2022minimizing} & Optimization & {\scriptsize\shortstack{various\\single-aisle}} &  & \checkmark & \checkmark & \checkmark & \checkmark &  & PRECISE by MIP \\
\addlinespace
\citet{bachmat2023air} & Physics model & \multicolumn{2}{c}{generic} & \checkmark &  &  & \checkmark &  & BF, slow-/fast-first, slow-back-first \\
\addlinespace
\citet{erland2024let} & Physics model & \multicolumn{2}{c}{generic} & \checkmark & \checkmark &  & \checkmark &  & slow-/fast-first \\
\addlinespace
\citet{jaehn2025determining} & Optimization & (3+3)*30 &  & \checkmark &  &  &  &  & BF, OI, RP, group policy by constraint programming \\
\midrule
\textbf{This paper} & \textbf{Deep RL} & \textbf{\checkmark} & \textbf{\checkmark} & \textbf{\checkmark} & \textbf{\checkmark} & \textbf{\checkmark} & \textbf{\checkmark} & \textbf{\checkmark} & \textbf{Learned dynamic policy} \\
\bottomrule
\end{longtable}

{\footnotesize\textsuperscript{a}\,Layout gives the nominal economy-class seating per row over $R$ rows (e.g.\ $(3{+}3)$: single aisle, three seats per side; $(3{+}4{+}3)$: double aisle). A few studies use layouts that deviate slightly, e.g.\ a shorter first or last row, or a small separate first-class section.}

{\footnotesize\textsuperscript{b}\,A checkmark indicates that the factor is explicitly used in the assignment decision. For example, companion groups are marked only when the method assigns companions together, not when companions are merely simulated under a companion-compatible policy.}

{\footnotesize\emph{Policy abbreviations:} BF, back-to-front; FB, front-to-back; AB, alternating-block; OI, outside-in (i.e., window--middle--aisle); SG, seat-group; RP, reverse-pyramid; PRECISE, per-passenger sequencing that assigns each passenger an exact position in the boarding queue; MILP, mixed-integer linear programming; MINLP, mixed-integer nonlinear programming; MIP, mixed-integer programming; OTHERS, policies not falling into the above categories.}
}
\end{landscape}

We develop a dynamic policy that does all three at once. We formulate the boarding group assignment as a Markov decision process (MDP): as each passenger checks in, the airline observes the seats and group assignments of previously checked-in passengers, together with the current passenger's seat and any luggage declaration, and assigns the boarding group, with travel companions always sharing a group. We parameterize the policy with a convolutional neural network (CNN) that reads the cabin as a grid, and optimize it by reinforcement learning (RL), specifically proximal policy optimization (PPO). To our knowledge, this is the first time the boarding problem has been formulated as an MDP and solved by RL.\par

We evaluate the dynamic policy against the static policies in a simulation model that captures aisle and seat interference, stochastic luggage handling, and passenger movement, across six representative single- and double-aisle commercial layouts, beyond the single-aisle focus of most prior work (Table~\ref{table:literature_review}). Unlike most studies, which target total boarding time alone, we optimize total and average individual boarding time jointly. Among the static policies, back-to-front with optimized group sizes is the strongest on both metrics. The dynamic RL policy beats it on both in every layout we test, and further outperforms the slow-first and fast-first luggage-ordering benchmarks \citep{bachmat2023air,erland2024let}. On a representative narrow-body airplane, the four-group RL frontier improves on the four-group optimal back-to-front benchmark by up to 9.8\% in total boarding time and up to 22.8\% in average individual boarding time, at opposite ends of the tradeoff between them. Sweeping a weight on the two metrics produces an approximate Pareto frontier, allowing the operator to select a preferred balance between operational efficiency and passenger experience. It also generalizes: a single trained policy continues to outperform the static policies under load factors, travel-companion distributions, and luggage distributions far from those used in training.\par

The remainder of this paper is organized as follows. Section~\ref{sec:literature_review} reviews the related literature on boarding policies and research methods. Section~\ref{sec:dynamic_group_problem} formulates the boarding group assignment as an MDP and distinguishes the static and dynamic policies evaluated within this framework. Section~\ref{sec:boarding_process} models the airplane boarding process, including detailed descriptions of various airplane layouts and passenger movement dynamics. Section~\ref{sec:static_boarding_policies} describes the three static policies: back-to-front, modified-Steffen, and alternating-block. Section~\ref{sec:rl_method} develops the dynamic RL policy, including its CNN-based state embedding, the actor-critic networks, and the PPO training algorithm. Section~\ref{sec:rl_evaluation} reports the numerical experiments: the static benchmarks, the Pareto tradeoff achieved by the RL policy, and robustness tests under shifted operating conditions. Section~\ref{sec:conclusions} summarizes the findings and discusses future work. Appendix~\ref{apdx:notation} lists the notations used in this paper.

\section{Literature Review}\label{sec:literature_review}

This section reviews boarding policies proposed to improve boarding efficiency across the academic literature and industry practice, followed by an examination of the research methods used to study the boarding process. Table~\ref{table:literature_review} summarizes the key works along the following dimensions: methodology (simulation, optimization, physics model, and field experiment), airplane layout (single- or double-aisle), the interference types modeled (aisle and seat), whether travel companions can board together, whether luggage information is used in the assignment decision, and the boarding policies evaluated. Across these works, the group assignment is always fixed in advance; none determines it dynamically as passengers check in, which is the direction this paper pursues.

\subsection{Boarding Policies}\label{sec:review_policies}

A boarding policy divides passengers into distinct groups and specifies the sequence for these groups to board the airplane; passengers within each boarding group may board in any order \citep{jaehn2015airplane}. At its most granular, a boarding policy could assign each passenger to an individual group, thereby determining a specific position in the boarding sequence (see e.g., \citealp{van2002reducing,steffen2008optimal,steffen2012experimental,willamowski2022minimizing}); related seating-assignment models also operate at seat-level resolution by assigning passengers to seats under a prescribed boarding order \citep{milne2016optimization,salari2019airplane}. This precise sequencing enables developing boarding orders that minimize the total boarding time \citep{jaehn2015airplane}.\par

However, real-world implementation of precise sequencing is difficult because passengers cannot reliably follow precise timing and positioning requirements in the queue, may arrive late, or need to board together with their companions. In practice, airlines usually rely on a limited number of boarding groups, substantially fewer than the total passenger count, to support gate management. Many seat-based boarding policies use seat assignments as their primary criterion and fall into three main categories: row-based, column-based, or a combination of both.\par

Row-based policies merit particular attention because they accommodate travel companions who board together. The most widely adopted row-based approach is the \textit{back-to-front} policy, which divides passengers into blocks according to row numbers and then passengers seated in the rear blocks board first, followed by those seated progressively toward the front \citep{van2005america,steffen2008optimal,bazargan2007linear,bachmat2013optimal}. The back-to-front policy has been adopted by carriers such as Alaska Airlines, and was used by Delta Air Lines during the COVID-19 pandemic. Some Chinese airlines like Sichuan Airlines have recently started implementing lighter versions of back-to-front boarding, by offering rear-seated passengers a 5-minute priority boarding window before the general boarding process. The \textit{front-to-back} policy operates in reverse, where passengers in the front rows board earlier than those in the back rows \citep{bachmat2009analysis,budesca2014optimization}. A more nuanced approach, the \textit{alternating-block} policy, segments the airplane into distinct row-based blocks but arranges them in an alternating sequence rather than following a strict directional order \citep{van2002reducing}. This policy was once used by the airline Ryanair.\par

Column-based approaches could potentially mitigate the seat interference but separate travel companions during boarding. The \textit{outside-in} policy exemplifies this approach, as previously implemented by United Airlines. This policy separates passengers into three sequential groups: window seat passengers board first, followed by middle seat passengers, then aisle seat passengers. This arrangement aims to eliminate the seat interference.\par

Other policies combine row and column information. The \textit{seat-group} policy merges outside-in and back-to-front, with each outside-in group further divided into blocks that board from back to front \citep{ferrari2005robustness,giitsidis2016modeling}. In the \textit{reverse-pyramid} policy, boarding groups form diagonal waves that move from back to front and from window to aisle simultaneously, which smooths the transition between groups. America West Airlines reduced boarding times by about 20\% with this policy \citep{van2005america}.

\subsection{Research Methods}

The airplane boarding problem can be viewed as a queuing problem. Classical analytical tools for queueing and traffic-service systems, including Markovian queueing models \citep{gu2011isolated,gu2015curbside,pacheco2017queues}, renewal or regenerative-cycle arguments \citep{heidemann1997queueing,gu2012highway,gu2013maximizing}, and approximation or bounding methods \citep{newell1982applications,shen2019capacity}, are difficult to apply directly to this problem. The difficulty stems from the diversity of airplane layouts, each with its own seating configuration, and from the stochastic and dynamic blockages and interferences between boarding passengers. The large passenger count, often exceeding 100 on commercial flights, together with the many possible boarding group configurations, further expands the solution space. Previous works therefore turned to other approaches, which fall into four categories: simulation, mathematical optimization, physics models, and field experiments.\par

Simulation has been widely used to examine queueing systems (e.g., \citealp{yang2020achieving,hu2023impacts,shen2023efficient}). It also stands as the predominant method in analyzing boarding processes. The approaches include agent-based models \citep{audenaert2009multi,coppens2018improving,budesca2014optimization,delcea2018investigating}, stochastic discrete event simulations \citep{van2002reducing,schultz2018field,schultz2018implementation}, cellular automata models \citep{giitsidis2016modeling}, and cellular discrete event simulation \citep{jafer2017comparative}. These methods examine various aspects such as passenger characteristics \citep{tang2012aircraft,ferrari2005improving}, infrastructure elements like pre-boarding areas \citep{steiner2009speeding} and apron bus operations \citep{milne2019new,cotfas2019testing}, and disruption scenarios \citep{delcea2018investigating}. Related work also examines how group behavior affects passenger motion, seat conflict, luggage handling, and boarding time \citep{tang2018aircraft}. \citet{zeineddine2017dynamically} proposes a dynamically optimized boarding queue that keeps traveling companions together, but the queue is generated only after the last passenger has checked in. Simulations excel at incorporating realistic passenger behaviors, aisle and seat interference patterns, and various operational constraints with greater flexibility. However, they face computational challenges, particularly when modeling large airplanes or conducting extensive policy comparisons, making it difficult to explore a wide range of operating conditions efficiently.\par

Mathematical optimization uses exact models to search over either seat-level or group-level decisions. \citet{milne2016optimization}, \citet{milne2018robust}, and \citet{salari2019airplane} optimize seating assignments, mainly using carry-on luggage information and then applying a prescribed boarding order, while \citet{willamowski2022minimizing} formulates total boarding time minimization directly and considers passenger-specific boarding positions. These models use mathematical programming formulations that can incorporate additional operational constraints, including companion constraints in some cases. Because passenger-level sequencing is difficult to enforce in practice, a second stream \citep{van2005america,bazargan2007linear,soolaki2012new,jaehn2025determining} optimizes assignments to a limited number of boarding groups rather than assigning every passenger an individual boarding position. Group-level optimization is less flexible than passenger-level sequencing or seating optimization, but it is more compatible with airport boarding operations.\par

Physics models, pioneered by Bachmat and colleagues \citep{bachmat2006analysis,bachmat2008bounds,bachmat2009analysis,bachmat2013optimal,bachmat2023air} and extended by Erland and colleagues \citep{erland2019lorentzian,erland2021lorentzian,erland2024let}, interpret airplane boarding through the lens of Lorentzian geometry --- the same mathematical framework used in relativity theory. These models represent boarding as wave propagation in a two-dimensional spacetime domain defined by queue position and row designation, where the boarding time corresponds to the longest causal path through this domain. The approach provides analytical formulas for boarding times as the number of passengers approaches infinity and is computationally efficient. The models also yield general insights, such as how boarding time scales with congestion level \citep{bachmat2013optimal} and why slow-first boarding universally outperforms fast-first under the Lorentzian framework \citep{erland2024let}. However, analytical tractability demands model simplifications: passengers are assumed to move at infinite speed when unobstructed, and the boarding time formulas are asymptotic estimates that become exact only as the number of passengers tends to infinity. These simplifications amplify the differences in boarding time between policies relative to more realistic simulations that account for finite walking speed \citep{erland2024let}.

Field experiments complement theoretical and simulation studies with empirical data. \citet{steffen2012experimental} conducted controlled experiments comparing five different boarding strategies with a mock airplane interior. \citet{schultz2018field} used field measurements to validate stochastic boarding models, achieving calibration errors within 2\%. Various studies examined specific boarding aspects: \citet{kierzkowski2017human} observed human factors and passenger behavior patterns, \citet{coppens2018improving} tested hand luggage stowing methods, and \citet{ren2018experimental} measured interference classifications during boarding. \citet{vendel2019effects} and \citet{hiemstra2019identifying} explored potential improvements through a hand luggage guiding system and passenger experience modifications respectively. \citet{gwynne2018small} documented detailed passenger micro-behaviors during boarding and deplaning. Field experiments provide empirical validation of theoretical models and identify practical operational constraints, but are limited by site-specific data collection, making them applicable only to a narrow range of operating conditions.

\section{Sequential Boarding Group Assignment Problem}\label{sec:dynamic_group_problem}

A boarding process involves two stages: passengers first select their seats and receive boarding group assignments during check-in, and then board the airplane in the designated group order. Check-in, whether at a physical counter or through a mobile application, is a natural point for implementing group-based boarding policies, as the assigned group number can be printed on or embedded in the boarding pass for clear reference at the boarding gate.

In this section, we formulate the group assignment task as a sequential decision problem and cast it as a Markov Decision Process (MDP). The formulation is policy-agnostic: any deterministic function that maps the operator's observation at check-in to a group assignment is a feasible policy. Both the existing static policies from the literature and the dynamic policy we develop in this paper can be evaluated within the same MDP-based decision framework; they differ in which state components they use and in whether the mapping is hand-designed or learned.

\subsection{Problem Description}
Passengers arrive sequentially at the check-in counter. Let $J$ denote the complete set of passengers and index them by arrival order, so that passenger $j+1$ arrives after passenger $j$. Each passenger selects a seat (row and column) and is then assigned to one of $N$ boarding groups by the airline operator. Within each group, passengers board in random order. The passenger set $J$ need not fill every seat; the formulation and all policies apply at any load factor.\par

The assignment decision for passenger $j$ depends on four factors: the selected seat, optional luggage declaration, travel companion group, and the check-in status of all previous passengers $\{1,\ldots,j-1\}$. The inclusion of luggage information is motivated by recent findings that the boarding order of passengers with different luggage-handling times affects both total boarding time and average individual boarding time, but in opposite directions \citep{bachmat2023air,erland2024let}. Although actual handling time is unknown before boarding, it correlates positively with the number of luggage items \citep{steiner2009speeding,erland2024let}. We therefore allow passengers to optionally declare their carry-on luggage quantity during check-in, and investigate how this information can be exploited by the assignment policy.\par

\begin{figure}[!htb]
    \centering
    \includegraphics[width=0.95\textwidth]{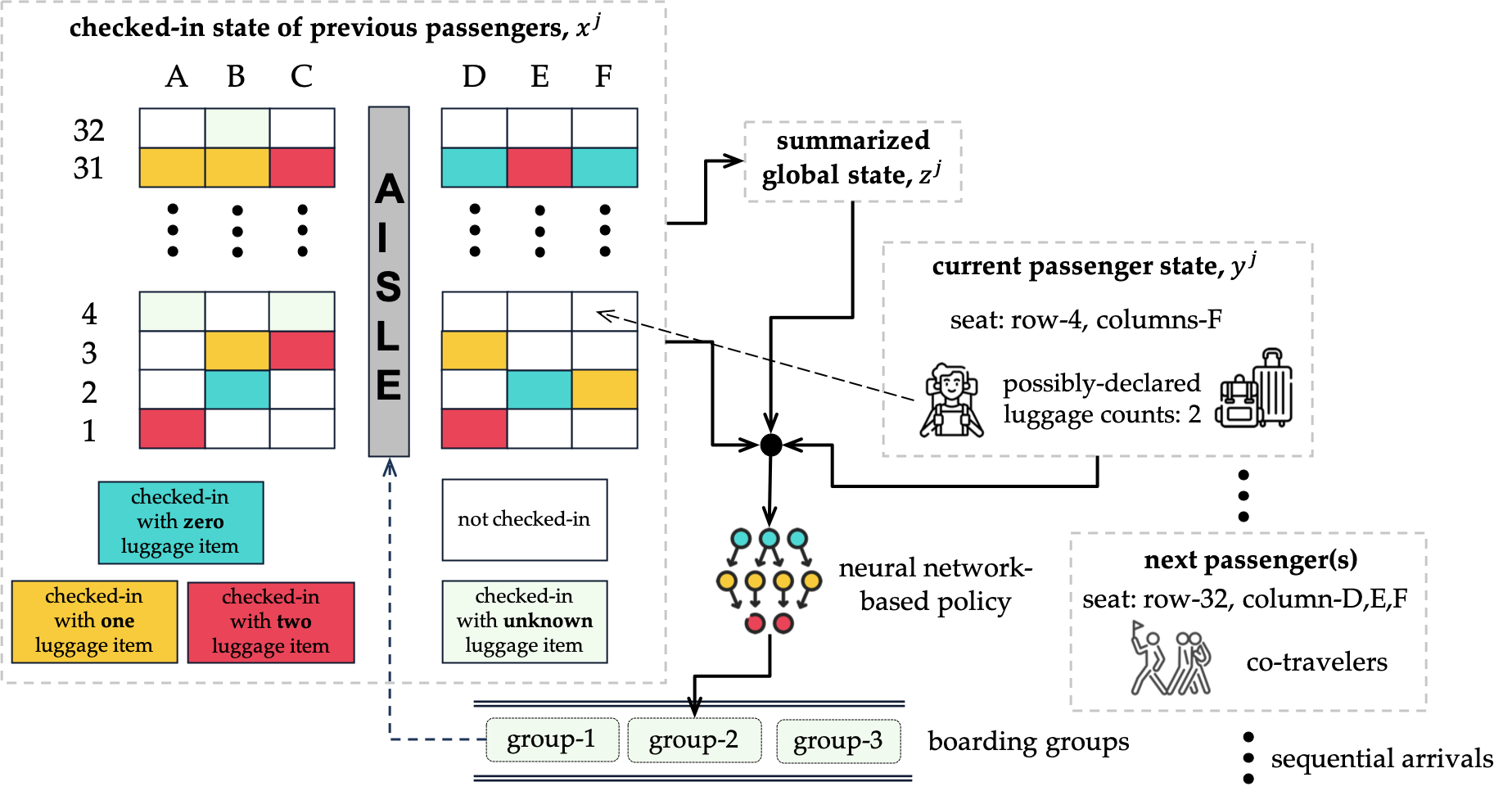}
    \caption{Illustration of the sequential boarding group assignment problem.}
    \label{fig:dynamic_assignment}
\end{figure}

Regarding travel companions, passengers who sit in adjacent seats may self-identify as a companion group during check-in and request to board together. All members of a companion group receive the same boarding group assignment. We model companion-group sizes up to a maximum of $M$ travelers with a distribution $(q_1,q_2,\ldots,q_M)$, where $q_k$ is the proportion of passengers traveling in groups of size $k$ and $\sum_{k=1}^{M}q_k=1$; in particular, $q_1$ denotes solo passengers. Fig.~\ref{fig:dynamic_assignment} summarizes the assignment problem and its input information.\par

\subsection{Markov Decision Process Formulation}\label{sec:mdp_formulation}

We formulate the above sequential group assignment problem as a $|J|$-step finite-horizon MDP, where $|J|$ denotes the total number of boarding passengers. The MDP consists of a 4-tuple $(\mathcal{S},\mathcal{A},\mathcal{P},\mathcal{R})$, where $\mathcal{S}$ represents the state space, $\mathcal{A}$ denotes the action space, $\mathcal{P}$ specifies the state transition dynamics, and $\mathcal{R}$ defines the reward function. The following sections detail each MDP element.

\subsubsection{State}\label{sec:mdp_state}

When the $j$-th passenger selects a seat, the agent (i.e., airline operator) observes a state $s^j \in \mathcal{S}$ consisting of three components:
\begin{enumerate}[itemsep=0pt, parsep=0pt, topsep=0pt, partopsep=0pt]
	\item \textbf{Checked-in state of previous passengers, $x^j$}: We represent the cabin layout as an $R \times C$ grid, where $R$ represents the number of rows and $C$ the number of columns, one entry per seat. The checked-in passengers $\{1,2,\ldots,j-1\}$ form a tensor $x^j \in \mathbb{R}^{R \times C \times D}$, where $D$ is the per-seat feature dimension. Each seat carries a $D$-dimensional feature vector with the following components:
		\begin{itemize}[itemsep=0pt, parsep=0pt, topsep=0pt, partopsep=0pt]
		\item An $(N+1)$-dimensional one-hot encoding vector where the first $N$ dimensions correspond to boarding groups $1$ through $N$, and the $(N+1)$-th dimension indicates an unselected seat;
		\item A 3-dimensional one-hot encoding vector specifying the seat location (window, middle, or aisle);
		\item The number of declared luggage items (if available) for the checked-in passenger.
	\end{itemize}

	The feature dimension $D$ therefore equals $(N+1)+(3)+(1) = N+5$.
	
	\item \textbf{Current passenger state, $y^j$}: A vector $y^j \in \mathbb{R}^7$ that encodes the passenger $j$'s selected row and column, the declared luggage count (if any), a 3-dimensional one-hot encoding vector for seat location (window, middle, or aisle), and the size of the passenger's companion group.
	
	\item \textbf{Global state, $z^j$}: A vector $z^j \in \mathbb{R}^{N+1}$ that records the total number of checked-in passengers and the per-group count.
\end{enumerate}

For undeclared luggage quantities, we use $-1$ as a default value.\footnote{While a one-hot encoding might enhance representation capability, our current treatment delivers satisfactory performance. We defer investigation of advanced encoding schemes to future work.}

The complete state observation is:
\begin{equation}
	s^j \equiv (x^j, y^j, z^j), \; x^j\in \mathbb{R}^{R \times C \times (N+5)}, \; y^j\in \mathbb{R}^7, \; z^j\in \mathbb{R}^{N+1}.
\end{equation}

\subsubsection{Action}

Given state $s^j$, a policy $\pi$ selects a boarding group $a^j=\pi(s^{j}) \in \mathcal{A}\equiv \{1,2,\ldots,N\}$. The action space is therefore small and discrete. When passenger $j$ belongs to a companion group, only the first member receives a policy-computed assignment; all subsequent members of the same group inherit this assignment, so that travel companions always share a boarding group.

\subsubsection{Transition Dynamics}

Once the group is assigned, the process advances to the next check-in decision, where a new state $s^{j+1}$ is observed. This system state evolution is characterized by a transition probability $\mathcal{P}\left( s^{j+1}| s^{j},a^j \right)$. The randomness in this process stems from the uncertainty regarding the next passenger's seat location and luggage count. Passenger $j+1$ is then assigned to a group determined by $a^{j+1} = \pi(s^{j+1})$, and this process continues for each subsequent passenger. Note that each state $s^{j}$ encapsulates previous allocation decisions, making this decision process a sequential decision problem where past actions influence future states.

\subsubsection{Reward}

Intermediate transitions yield zero reward: $r^j = 0$ for $j = 1,\ldots,|J|-1$. After the final assignment has been made, the actual boarding process commences and the boarding outcome is revealed. We consider two performance metrics: total boarding time $T_{\mathrm{total}}$, measured from the first passenger's entry until the last passenger is seated, and average individual boarding time $T_{\mathrm{avg}}$, defined as the mean time each passenger spends from the start of boarding to being seated. \citet{bachmat2023air} show that under different luggage-based ordering rules (slow-first and fast-first), these two metrics can move in opposite directions. To account for this potential tradeoff, we combine both metrics in the terminal reward:
\begin{equation}\label{eq:two_obj}
r^{|J|} =
-\left[
(1-\lambda)\frac{T_{\mathrm{total}}}{T_{\mathrm{total}}^{\mathrm{rand}}}
+
\lambda\frac{T_{\mathrm{avg}}}{T_{\mathrm{avg}}^{\mathrm{rand}}}
\right],
\end{equation}
where $T_{\mathrm{total}}^{\mathrm{rand}}$ and $T_{\mathrm{avg}}^{\mathrm{rand}}$ are the corresponding mean metrics under random boarding for the same layout and operating setting. The parameter $\lambda\in[0,1]$ controls the tradeoff, with $\lambda=0$ targeting normalized total boarding time alone and $\lambda=1$ targeting normalized average individual boarding time alone. This normalization makes the two terms comparable in scale.

\subsubsection{Objective}
The goal is to find a policy $\pi$ that maximizes the expected episodic return:
\begin{equation}
\max_{\pi}\; \mathbb{E}[G] = \mathbb{E}\left[\sum_{j=1}^{|J|} r^j\right],
\end{equation}
where $G$ sums rewards over all $|J|$ passengers in an episode. Since only the terminal reward is nonzero, maximizing $\mathbb{E}[G]$ minimizes the normalized weighted combination in Eq.~\eqref{eq:two_obj}. This sparse-reward structure is standard in finite-horizon MDPs; each group assignment alters the checked-in assignment state observed by subsequent decisions and thus affects the terminal boarding outcome.

\subsection{Two Policy Classes}\label{sec:policy_classes}

The MDP formulated above is agnostic to the form of the policy: any deterministic function $\pi: \mathcal{S} \rightarrow \mathcal{A}$ that respects the co-traveler constraint in Section~\ref{sec:mdp_formulation} is a candidate solution. In the remainder of the paper we examine two distinct policy classes within this MDP-based framework.\par

\textbf{Static policies.} A static policy follows a fixed, predetermined rule that maps seat location to a group label, without observing the check-in history or adapting to the evolving checked-in assignment state. Depending on the design, such a rule may or may not guarantee that co-travelers in adjacent seats board within the same group. The three static policies examined in this paper (back-to-front, modified-Steffen, and alternating-block, presented in Section~\ref{sec:static_boarding_policies}) all keep co-travelers together.\par

\textbf{Dynamic policies.} In contrast, a dynamic policy operates on the full state $s^j = (x^j, y^j, z^j)$ and adapts its group assignment to the evolving checked-in assignment state as each passenger checks in. We develop such a policy in Section~\ref{sec:rl_method}, parameterized by deep neural networks and trained via RL.

\section{Modeling Airplane Boarding Process}\label{sec:boarding_process}

This section details a simulation model that, given any sequence of boarding-group assignments produced by a policy on the MDP of Section~\ref{sec:dynamic_group_problem}, generates the total boarding time $T_{\mathrm{total}}$ and average individual boarding time $T_{\mathrm{avg}}$ used in the terminal reward. Section~\ref{sec:airplane_layout} specifies the full range of airplane layouts studied, Section~\ref{sec:passenger_kinetics} models passenger kinetics in the aisle and at the seat, and Section~\ref{sec:simulation} describes the simulation implementation.

\subsection{Airplane Layouts}\label{sec:airplane_layout}

\begin{figure}[!htb]
    \centering
    \includegraphics[width=0.9\textwidth]{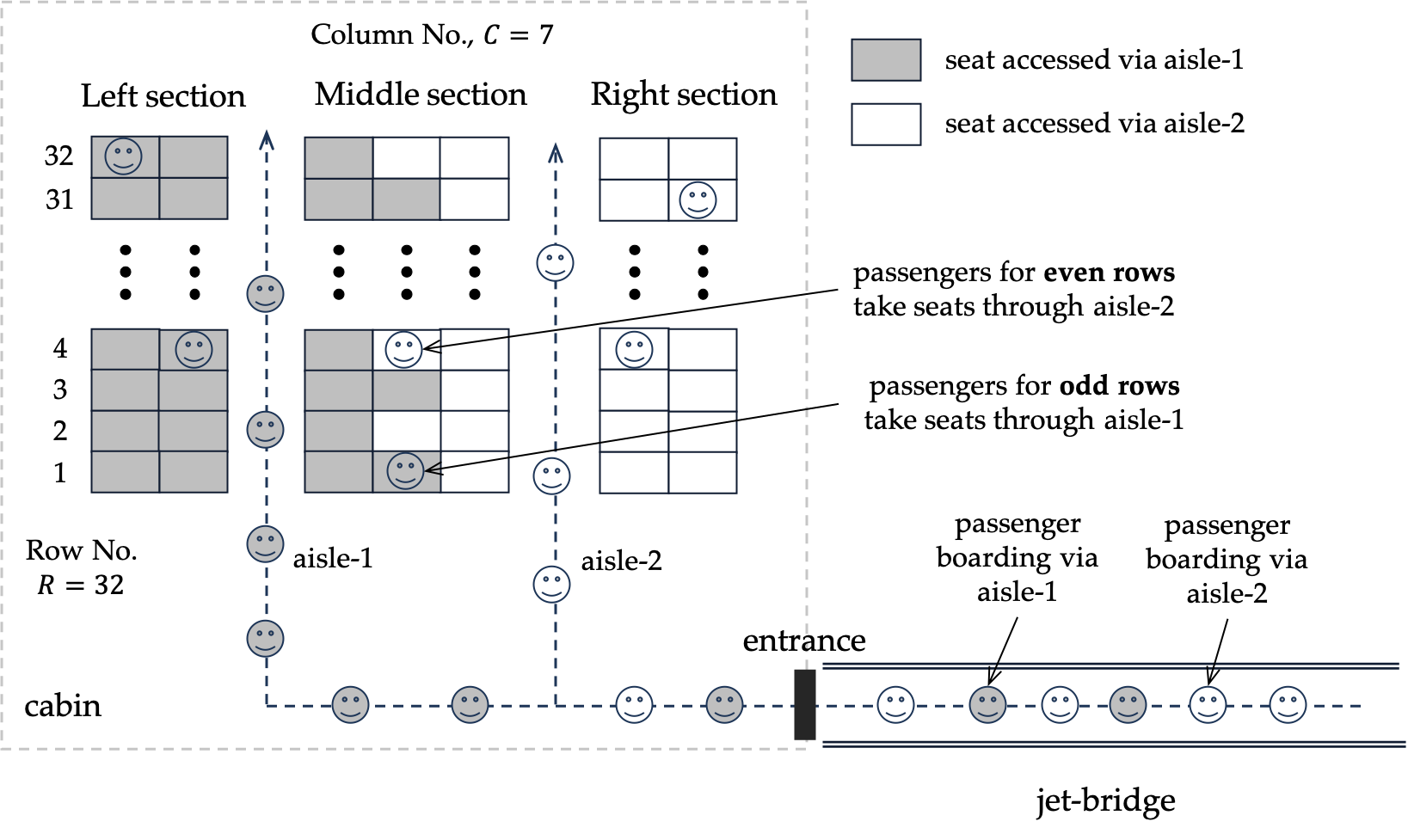}
    \caption{Schematic representation of a double-aisle airplane layout and its boarding process.}
    \label{fig:layout_double_aisle}
\end{figure}

We focus on the most prevalent boarding setting found in real-world practice. This setting involves a single entrance at the front of the passenger cabin, typically accessed via a jet-bridge \citep{giitsidis2016modeling, willamowski2022minimizing}. All passengers form a single queue before this entrance prior to entering the cabin \citep{zeineddine2017dynamically}. The cabin itself consists of either one or two aisles with seats on both sides \citep{airbus2024gmf}. In the case of a double-aisle layout, the queue bifurcates as passengers enter the cabin, with individuals proceeding to their respective aisles (see Fig.~\ref{fig:layout_double_aisle} for an illustration of the double-aisle case). Data from \citet{airbus2024gmf} indicate that wide-body airplanes with double-aisle layouts constitute 21\% of the total global airplane fleet.

Our study employs the commonly-used ``economy-class boarding'' assumption, focusing on optimizing the boarding process for economy class passengers, as they constitute the majority of passengers and have the most significant impact on overall boarding times. This assumption excludes priority boarding (e.g., first class) and other pre-boarding procedures from our analysis \citep{steffen2008optimal,willamowski2022minimizing}, since they have fewer passengers and are unlikely to substantially affect the boarding performance.\par

We classify representative economy-class cabin layouts based on the number of seats per row and the presence of either a single aisle or double aisles. We denote the total number of rows as $R$ and the number of seats per row (column number) as $C$. Our study examines the following economy-class cabin layouts (data sourced from \citealp{aircraftdb2024}):

\begin{enumerate}[itemsep=0pt, parsep=0pt, topsep=0pt, partopsep=0pt]
	\item $(2+2)*R$ ($C=4$): A single-aisle layout with two seats on each side. This layout appears in smaller airplanes like the Bombardier CRJ series.
	\item $(3+3)*R$ ($C=6$): A single-aisle layout with three seats on each side. This layout characterizes popular narrow-body airplanes, such as the Airbus A320 and A321, and Boeing 737 series.
	\item $(2+3+2)*R$ ($C=7$): A double-aisle layout with seven seats per row. The middle section contains three seats, while each side section has two seats. This layout typifies wide-body airplanes like the Boeing 767 series.
	\item $(2+4+2)*R$ ($C=8$): A double-aisle layout with eight seats per row. The middle section contains four seats, while each side section has two seats. This layout appears in wide-body airplanes such as the Airbus A330 and A340 series.
	\item $(3+3+3)*R$ ($C=9$): A double-aisle layout with nine seats per row, evenly distributed across three sections. This layout exists in larger wide-body airplanes, including the Boeing 787 and Airbus A350 series.
	\item $(3+4+3)*R$ ($C=10$): A double-aisle layout with ten seats per row, representing the widest passenger planes currently in service. The Boeing 777 and Airbus A380 series exemplify this layout.
\end{enumerate}

These layouts span a broad set of widely used commercial airplane designs, from regional jets to large wide-body airplanes. By varying the number of rows ($R$), we can represent a wide range of economy-class cabin sizes. This flexibility in row numbers also reflects real-world variations, where budget airline companies may opt to increase seating density by adding more rows, while other carriers might choose a more spacious layout with fewer rows.

\subsection{Passenger Kinetics}\label{sec:passenger_kinetics}

To simulate passenger movements, we discretize time with step size $\Delta t=1.2$ seconds and divide the space along an aisle into cells. Each row occupies two cells. Passengers can move at most one cell per $\Delta t$. We consider the boarding process that spans from the moment the first passenger enters the airplane until the last passenger is seated.

\subsubsection{Forward Movement in Aisles}\label{sec:moving_forward}

\begin{figure}[!tbp]
    \centering
    \includegraphics[width=0.75\textwidth]{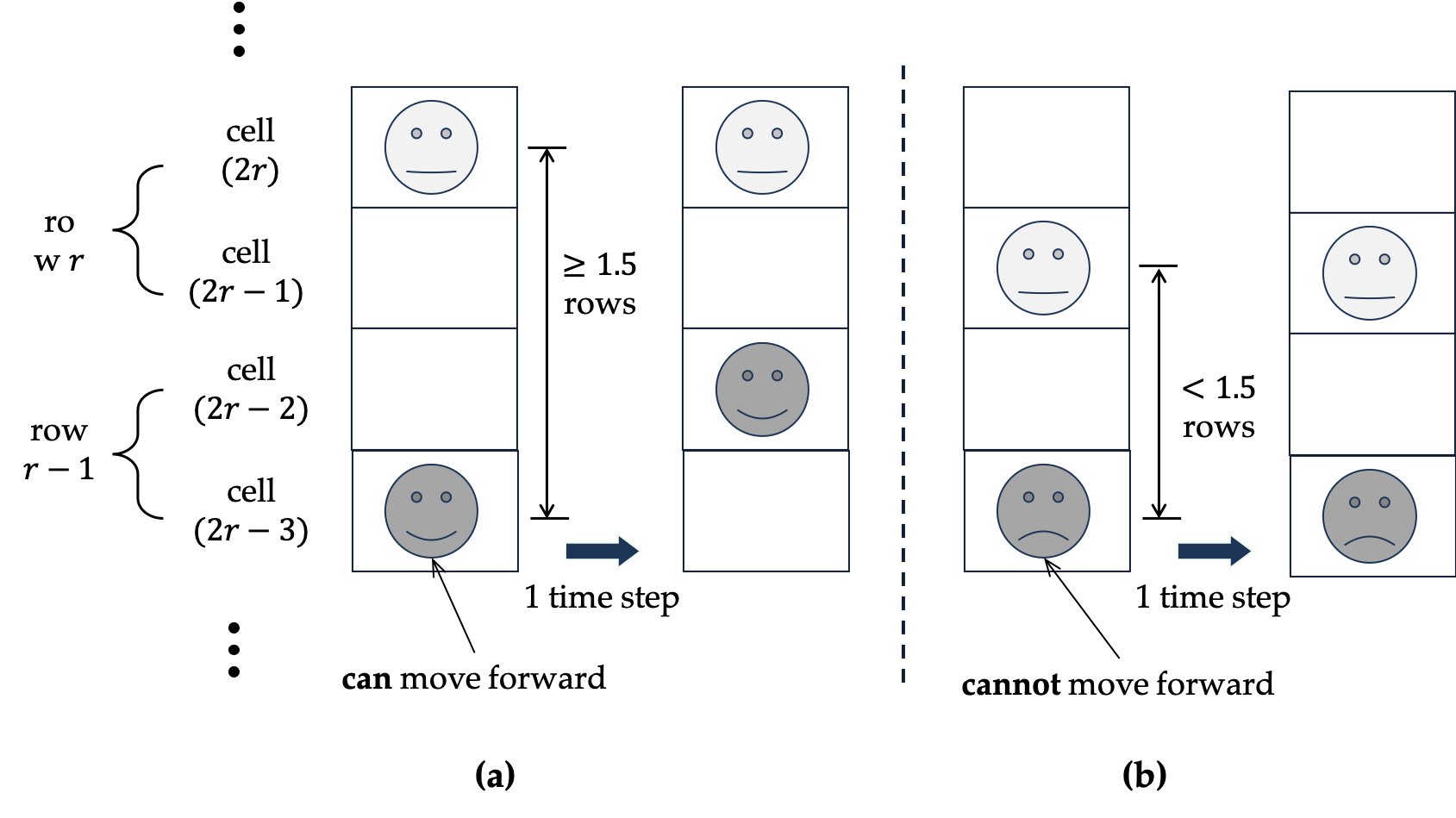}
    \caption{Illustration of passengers' forward movements in an aisle: (a) When the passenger spacing is $\geq 3$ cells, movement is allowed. (b) When the spacing is $< 3$ cells, the passenger stops.}
    \label{fig:moving_forward}
\end{figure}

The forward progression of each passenger depends on their spacing to the passenger ahead, consistent with grid-based boarding models in which aisle movement is forward-directed and constrained by local cell occupancy \citep{schultz2018implementation}. The movement rules follow two distinct scenarios:
\begin{enumerate}[itemsep=0pt, parsep=0pt, topsep=0pt, partopsep=0pt]
	\item A passenger can advance one cell (equivalent to half a row) per time step if the distance to the passenger ahead equals or exceeds three cells (1.5 rows);
	\item A passenger must maintain their current position if the distance to the passenger ahead is less than three cells.
\end{enumerate}
Fig.~\ref{fig:moving_forward} provides a visual representation of these two movement scenarios.

\subsubsection{Luggage Stowing}\label{sec:luggage_stowing}

Upon reaching the destination row, a passenger stores luggage before entering their seat. The narrow cabin aisle forces this passenger to block the aisle during luggage handling, so other passengers behind cannot proceed to their assigned rows. Note that the spacing rules in Section~\ref{sec:moving_forward} allow enough space for a passenger to handle luggage. We model the luggage-handling time as follows:
\begin{enumerate}[itemsep=0pt, parsep=0pt, topsep=0pt, partopsep=0pt]
	\item The number of luggage items per passenger, $\boldsymbol{K}$, follows a categorical distribution on support $k=\{0, 1, 2, 3\}$, with probabilities $\{p_k\}_{k=0,1,2,3}$;
	\item For a given realization $k$ of $\boldsymbol{K}$, the total luggage-handling time, $\boldsymbol{O}_k$, follows a distribution with mean $\mu_k$ and standard deviation $\sigma_k$.
\end{enumerate}
A realized total luggage-handling time $L$ for a passenger is thus given by:
\begin{equation}
L = o_k, \quad k\sim \boldsymbol{K}, \quad o_k\sim \boldsymbol{O}_k.
\end{equation}

The choice of a categorical distribution for luggage items and the associated time distributions is based on observed passenger behaviors \citep{steiner2009speeding}. This formulation captures variability in both the number of items and the time to store them, and is flexible enough to represent the complex empirical distributions reported in the literature (e.g., \citealp{erland2024let}). As a special case, the constant luggage-handling time used in \citet{ferrari2005improving} can be modeled by setting $\boldsymbol{K}$ to a single value $k$ and $\sigma_k =0$.

\subsubsection{Seat Interferences}\label{sec:seat_interferences}
Seat taking occurs after the luggage storing process. Seat interference refers to the situation where, upon arrival at the designated row, an incoming passenger is blocked from reaching her designated seat by other passengers already seated in the aisle and middle seats of the row. This results in an extended aisle blockage.

\begin{figure}[!tbp]
    \centering
    \includegraphics[width=0.48\textwidth]{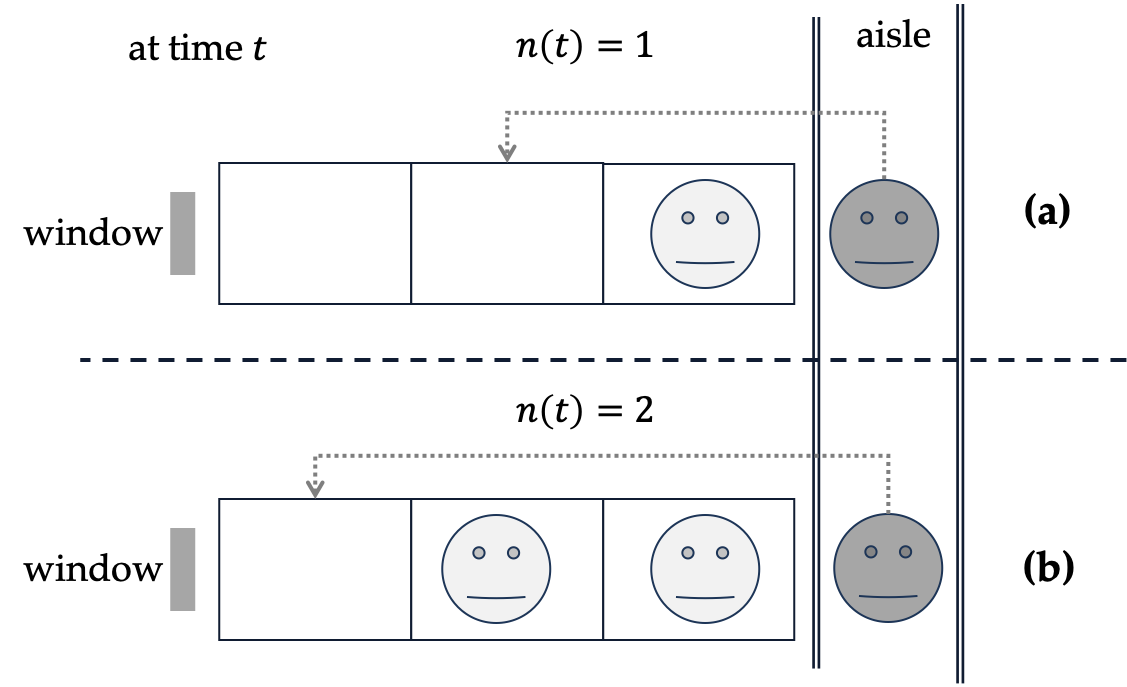}
    \caption{Examples of seat interference scenarios: (a) A passenger accessing a middle seat with the aisle seat occupied. (b) A passenger accessing a window seat with both aisle and middle seats occupied.}
    \label{fig:seat_interference}
\end{figure}

To account for seat interference, we model the seating time as a function of the number of occupied seats between a passenger and her seat \citep{ferrari2005robustness}:
\begin{equation}
S(n(t)) = (2n(t)+1)\cdot \xi,
\end{equation}
where $n(t)$ is the number of occupied seats between the aisle and the passenger's designated seat at the moment $t$ when she reaches her row, evaluated independently for each passenger, and $\xi$ represents the basic time unit required for a single passenger movement. The term $2n(t)$ accounts for the time needed for seated passengers to stand up and sit back, while the additional time unit (+1) represents the fixed time for the incoming passenger to move to her seat. Fig.~\ref{fig:seat_interference} illustrates two possible scenarios of determining the $n(t)$.

\subsubsection{Boarding Process of Double-Aisle Layouts}

The presence of two aisles divides the cabin into three sections: left, middle, and right; see again Fig.~\ref{fig:layout_double_aisle}. Passengers with seats in the left section use aisle-1, while those with seats in the right section use aisle-2. The middle section requires a more nuanced approach depending on its seat layout.

In layouts with four columns in the middle section, we assume that passengers seated in the two columns on the left will use aisle-1 for boarding, while those in the two right columns will use aisle-2.

For middle sections with three columns, we make the following assumption for simplicity of simulation (see Fig.~\ref{fig:layout_double_aisle}). Passengers seated in the leftmost column use aisle-1, and those in the rightmost column use aisle-2. For the central column, we alternate the aisle usage based on row numbers: passengers in odd-numbered rows use aisle-1, while those in even-numbered rows use aisle-2. This alternating pattern provides a balanced and tractable approximation for distributing the central-column passenger flow between the two aisles. In reality, passengers might choose either aisle based on various factors such as personal preference, perceived congestion, or airline instructions. While it is challenging to capture all these nuances, our alternating approach provides a balanced and tractable way to simulate passenger flows in these layouts.\footnote{We have conducted numerical experiments in which the passenger's aisle choice of the central column is randomized, and found that the findings presented in this paper are not affected.}

\subsection{Simulation}\label{sec:simulation}
       
We developed a parsimonious, discrete-time simulation program in-house using Python. It encapsulates the dynamics of passenger movements, luggage handling, and aisle and seat interferences described in the previous sections. All simulations were executed on a Mac Studio equipped with an M3 Ultra processor and 512 GB memory. The average runtime for one boarding simulation was approximately 0.1 seconds.

\section{Static Boarding Policies}\label{sec:static_boarding_policies}

This section presents three static boarding policies from the literature: back-to-front, modified-Steffen, and alternating-block. Each follows a fixed rule that maps a passenger's seat location to a boarding group, without observing the check-in history or checked-in assignment state. All three allow travel companions with adjacent seats to board together.\par

\subsection{Back-to-Front Policy}

\begin{figure}[!htb]
    \centering
    \includegraphics[width=0.55\textwidth]{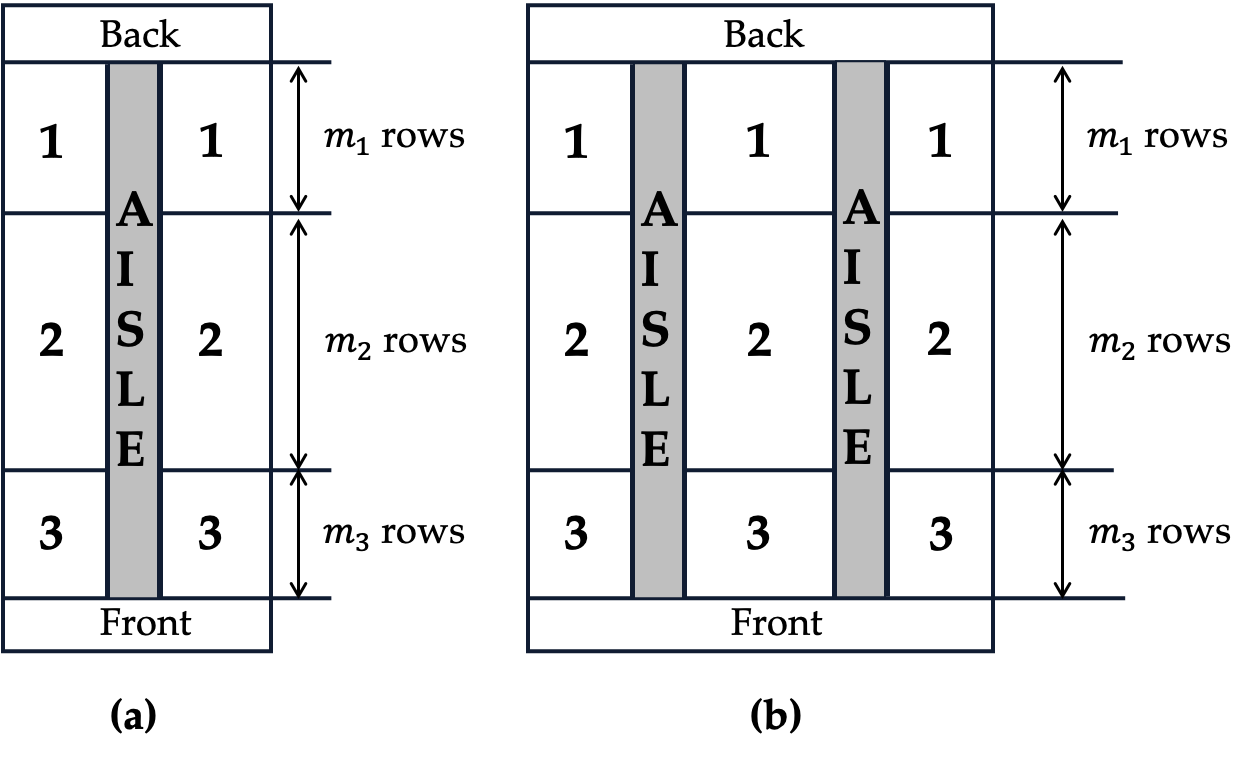}
    \caption{Back-to-front policy with 3 boarding groups: (a) Single-aisle layout; (b) Double-aisle layout.}
    \label{fig:bf_policy}
\end{figure}

The back-to-front policy categorizes passengers into $N$ groups of consecutive rows. The group combination is defined by an array $[m_1, m_2, \ldots, m_N]$, where the group number $i \in \{1,2,\ldots,N\}$ is indexed from back to front, and $m_i$ is the number of rows assigned to group $i$, with $\sum_{i=1}^{N}{m_i} = R$. Specifically, the boarding process occurs in stages: the first group consists of rows ${R, R-1, \ldots, R-m_1+1}$;
	the second group comprises rows ${R-m_1, R-m_1-1, \ldots, R-m_1-m_2+1}$; and so on. Within each group, passengers queue up in random order. Fig.~\ref{fig:bf_policy} illustrates a back-to-front plan with three groups.

Many previous studies have suggested that back-to-front boarding with multiple groups is less effective than random boarding \citep{ferrari2005improving,audenaert2009multi,bachmat2009analysis,milne2014new,jafer2017comparative,delcea2018investigating}. However, these studies generally assumed equal group sizes (i.e., $m_1 = m_2 = \ldots = m_N$). The elegant analytical model presented by \citet{bachmat2013optimal} was the first to challenge this assumption, showing that unequal group sizes can make back-to-front outperform random boarding. However, their model did not account for seat interferences or model double-aisle layouts precisely. Therefore, the exact performance of optimized back-to-front policies against other prevailing companion-compatible policies, especially under double-aisle layouts, remains an open question.

\subsection{Modified-Steffen Policy}

\begin{figure}[!htb]
    \centering
    \includegraphics[width=0.4\textwidth]{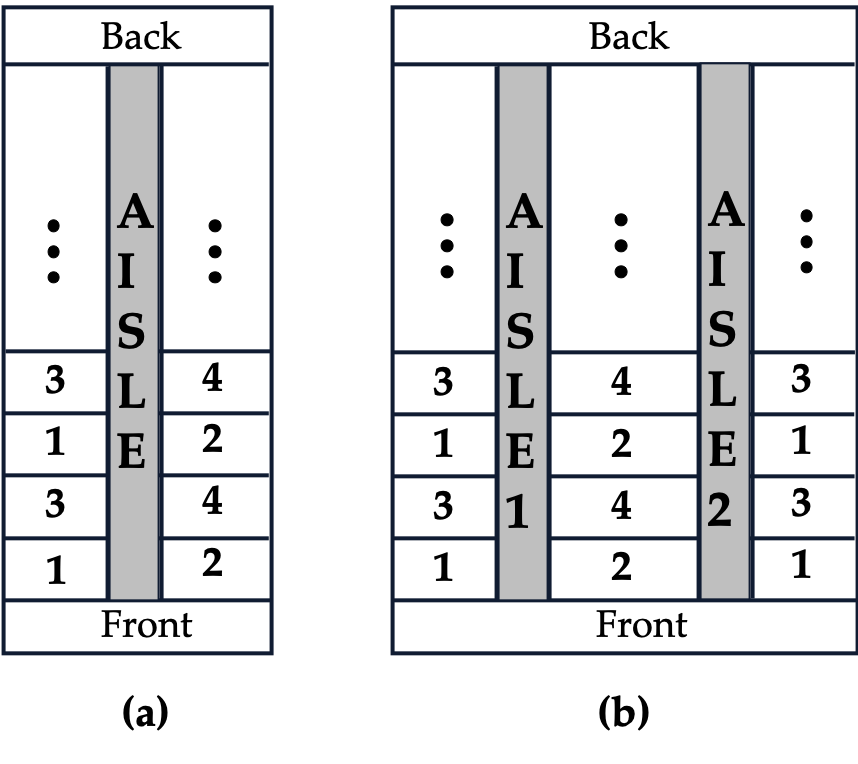}
    \caption{Modified-Steffen policy with 4 boarding groups: (a) Single-aisle layout; (b) Double-aisle layout.}
    \label{fig:ms_policy}
\end{figure}

The modified-Steffen policy uses four boarding groups, as illustrated by Fig.~\ref{fig:ms_policy}. The original modified-Steffen rule was developed for single-aisle airplanes \citep{steffen2008optimal}. We extend it to double-aisle layouts as follows. Recall from Section~\ref{sec:boarding_process} that a \textit{section} denotes the seats on one side of an aisle, so a single-aisle cabin has two sections and a double-aisle cabin has three (left, middle, and right). Within each section, the groups occupy alternating rows, so adjacent rows belong to different groups; on any given row, seats in adjacent sections also belong to different groups. Boarding proceeds in the order of groups 1, 2, 3, and 4, with random passenger queueing within each group.

\subsection{Alternating-Block Policy}

\begin{figure}[!htb]
    \centering
    \includegraphics[width=0.52\textwidth]{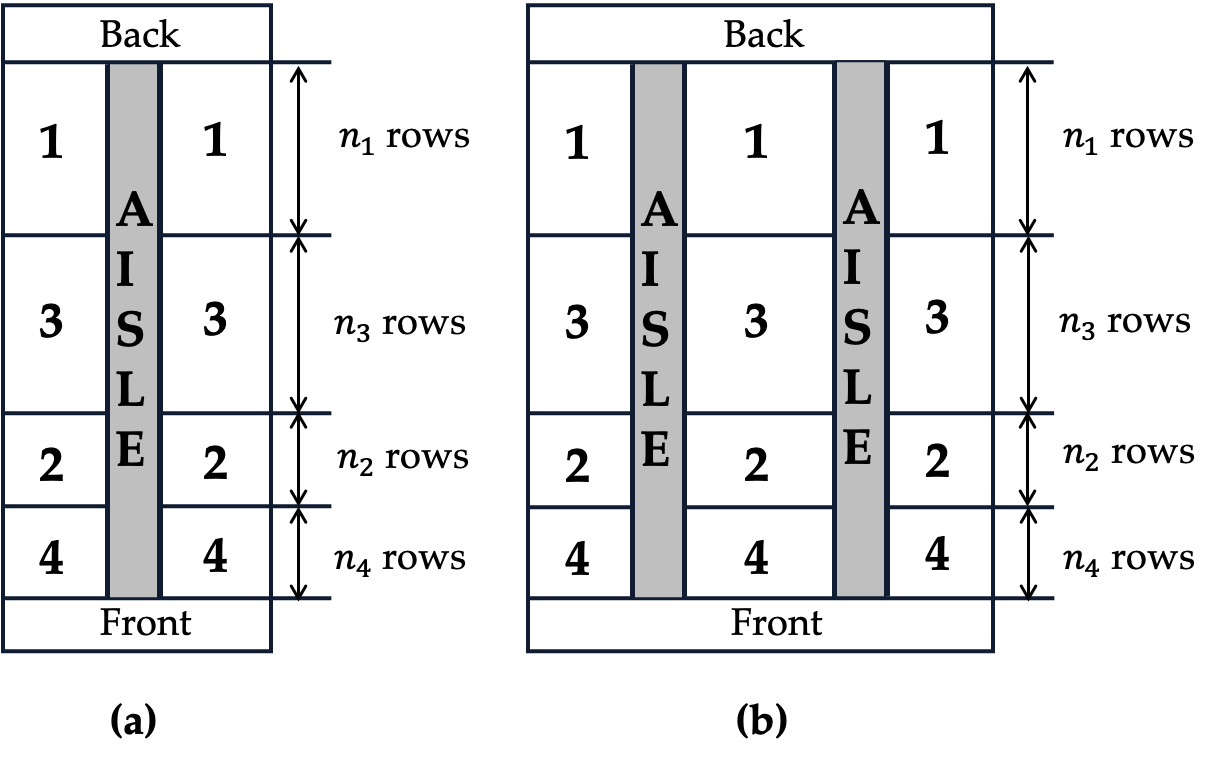}
    \caption{Alternating-block policy with 4 boarding groups. (a) Single-aisle layout; (b) Double-aisle layout.}
    \label{fig:alternating_block}
\end{figure}

The alternating-block policy segments the airplane into row-based blocks, each a set of consecutive rows. It boards these blocks in an alternating rather than sequential order (see Fig.~\ref{fig:alternating_block}). Alternating-block boarding in \citet{van2002reducing} can be implemented with different block sequences and numbers of blocks. We fix a four-block alternating order and optimize only the block sizes. This keeps the benchmark comparable to the four-group modified-Steffen policy while limiting operational complexity for airport staff. Group-1 passengers are assigned to the rearmost block, group-2 passengers to the third block from the rear, group-3 passengers to the second block from the rear, and group-4 passengers to the frontmost block. We index each block by the group assigned to it, so block $i$ contains group $i$ and comprises $n_i$ rows, $i\in\{1,2,3,4\}$.

This design intends to reduce the blockage between consecutive boarding groups, but it brings two problems in practice. First, a later group can be held up by the group just ahead of it in the boarding order. In the second phase, group-3 passengers sit behind group 2 and must walk through block 2 to reach their own. But group 2 might still be stowing luggage and taking seats in block 2 when group 3 is boarding. Second, a large leading group spills into the blocks ahead of it. A large group 1, for instance, may reach forward through block 3 into block 2 and interfere with group 2's boarding. Both effects undermine the parallel boarding the design assumes.\par

\section{Dynamic Boarding Policy via Deep Reinforcement Learning}\label{sec:rl_method}

We solve the MDP in Section~\ref{sec:mdp_formulation} with PPO \citep{schulman2017proximal} under the actor-critic framework. The actor outputs a probability distribution over boarding groups for each arriving passenger; the critic estimates the state-value function and provides the advantage baseline for policy updates. Section~\ref{sec:state_embedding} presents the state embedding, Section~\ref{sec:actor_critic_network} defines the actor and critic networks, and Section~\ref{sec:training_algorithm} details the training procedure.

\subsection{State Embedding}\label{sec:state_embedding}

The state $s^j=(x^j, y^j, z^j)$ defined in Section~\ref{sec:mdp_state} has three components that capture different aspects of the assignment problem: the cabin-wide checked-in assignment pattern ($x^j$), the current passenger's characteristics ($y^j$), and the aggregate group counts ($z^j$). We first embed each component through a dedicated neural network module and then concatenate the results into a unified state representation that serves as input to both the actor and critic networks in Section~\ref{sec:actor_critic_network}.

\subsubsection{Embedding of Checked-In Passenger States}\label{sec:embedding_checked_in}

The checked-in passenger state $x^j\in \mathbb{R}^{R\times C\times D}$ is passed through three 2D convolutional layers with activation function $\phi$: $\text{Conv2D}_1$ with $F_1$ filters, $\text{Conv2D}_2$ with $F_2$ filters, and $\text{Conv2D}_3$ with $F_3$ filters, all using $3\times 3$ kernels and padding~$=1$:
\begin{equation}
	h_1 = \phi(\text{Conv2D}_1(x^j)),\quad h_2 = \phi(\text{Conv2D}_2(h_1)), \quad h_3 = \phi(\text{Conv2D}_3(h_2)),
\end{equation}
%\begin{equation}
%	h_2 = \phi(\text{Conv2D}_2(h_1)),
%\end{equation}
%\begin{equation}
%	h_3 = \phi(\text{Conv2D}_3(h_2)),
%\end{equation}
where $h_1 \in \mathbb{R}^{R \times C\times F_1}$, $h_2 \in \mathbb{R}^{R \times C\times F_2}$, and $h_3 \in \mathbb{R}^{R \times C\times F_3}$.

To reduce spatial dimensions and capture global features, we then apply Global Average Pooling (GAP). The GAP operation computes the average of each feature map:
\begin{equation}
g_i = \frac{1}{R \cdot C} \sum_{k=1}^R \sum_{l=1}^C [h_3]_{k,l,i},
\end{equation}
where $g_i$ is the $i$-th element of the output vector $g\in \mathbb{R}^{F_3}$ and $[h_3]_{k,l,i}$ represents the element at position $(k,l,i)$ in the tensor $h_3$. The complete vector output by the GAP operation can be written as:
\begin{equation}
g = \left[\frac{1}{R \cdot C} \sum_{k=1}^R \sum_{l=1}^C [h_3]_{k,l,1} \; ,\ldots, \; \frac{1}{R \cdot C} \sum_{k=1}^R \sum_{l=1}^C [h_3]_{k,l,F_3}\right].
\end{equation}

Finally, a dense (i.e., fully connected) layer reduces the pooled features to a $G_x$-dimensional vector:
\begin{equation}
e_x = \phi(W_x\cdot g + b_x),
\end{equation}
where $W_x \in \mathbb{R}^{G_x \times F_3}$ and $b_x \in \mathbb{R}^{G_x}$ are the weight matrix and bias vector of the dense layer, respectively, and $e_x \in \mathbb{R}^{G_x}$ is the embedded representation of $x^j$.

\subsubsection{Embedding of Current Passenger State}\label{sec:embedding_current_passenger}

The current passenger state $y^j \in \mathbb{R}^7$ passes through a single dense layer:
\begin{equation}
e_y = \phi(W_y\cdot y^j + b_y),
\end{equation}
where $W_y \in \mathbb{R}^{G_y \times 7}$ and $b_y \in \mathbb{R}^{G_y}$ are the weight matrix and bias vector respectively, and $e_y \in \mathbb{R}^{G_y}$ is the resulting $G_y$-dimensional embedding of the passenger's characteristics.

\subsubsection{Embedding of Global State}\label{sec:embedding_global}
The global state $z^j \in \mathbb{R}^{N+1}$ is embedded using another single dense layer:
\begin{equation}
e_z = \phi(W_z\cdot z^j + b_z),
\end{equation}
where $W_z \in \mathbb{R}^{G_z \times (N+1)}$ and $b_z \in \mathbb{R}^{G_z}$ are the weight matrix and bias vector respectively, and $e_z \in \mathbb{R}^{G_z}$ is the resulting $G_z$-dimensional embedding of the global state.

\subsubsection{Final State Embedding}

The embeddings from the above three components (Sections \ref{sec:embedding_checked_in}-\ref{sec:embedding_global}) are concatenated and fed into a dense layer to create the final state representation with dimension $H$:
\begin{equation}
h^j = \phi(\bar{W}\cdot \bar{e} + \bar{b}),
\end{equation}
where $\bar{e} = [e_x; e_y; e_z] \in \mathbb{R}^{G_x+G_y+G_z}$, $\bar{W} \in \mathbb{R}^{H\times (G_x+G_y+G_z)}$, $\bar{b} \in \mathbb{R}^{H}$, and $h^j \in \mathbb{R}^{H}$.

In summary, we can express the entire embedding process $\Xi(\cdot)$ symbolically as:
\begin{equation}
	h^{j} = \Xi\left( s^{j}\right), \text{where}\; s^j \equiv (x^j\in \mathbb{R}^{R \times C \times (N+5)}, y^j\in \mathbb{R}^{7}, z^j\in \mathbb{R}^{N+1}), \; \text{and}\; h^j \in \mathbb{R}^{H}.
\end{equation}

\subsection{Actor and Critic Networks}\label{sec:actor_critic_network}

Both the actor and the critic use the embedding architecture from Section~\ref{sec:state_embedding}, but each maintains its own copy of the parameters, producing $h^j_\theta=\Xi_\theta(s^j)$ and $h^j_\omega=\Xi_\omega(s^j)$ respectively.\par

The actor's policy function $\pi_\theta$, parameterized by $\theta$, receives the state observation $s^{j}$ and produces the probability of selecting action $a^j$. The actor network begins with the aforementioned embeddings and processes them as follows:
\begin{equation}
    \pi_\theta(a^j|s^j)=\left[\text{Softmax}\left(W_{\text{actor}}\cdot h^j_\theta + b_{\text{actor}}\right)\right]_{a^j}, \quad a^j\in\{1,2,\ldots,N\},
\end{equation}
where $W_{\text{actor}}\in \mathbb{R}^{N\times H}$ and $b_{\text{actor}} \in \mathbb{R}^{N}$ are the weight matrix and bias vector of the dense layer, respectively. A softmax activation is applied to output a probability distribution over the boarding groups. Note that $\theta$ also includes the parameters used for state embedding.\par

The critic reduces the variance of policy-gradient estimates by providing a state-dependent baseline. It learns a state-value function $V_{\omega}$, parameterized by $\omega$, which takes the state observation $s^{j}$ as input and outputs an estimate $V_\omega(s^{j})$:
\begin{equation}
	V_\omega(s^{j})=W_{\text{critic}}\cdot h^j_\omega + b_{\text{critic}},
\end{equation}
where $W_{\text{critic}}\in \mathbb{R}^{1\times H}$ and $b_{\text{critic}} \in \mathbb{R}^{1}$ are the weight matrix and bias vector, respectively. 

\subsection{Training Algorithm}\label{sec:training_algorithm}

The training procedure collects complete check-in trajectories and updates the actor and critic using PPO. For episode $b$, let $\tau_b=\{(s_b^j,a_b^j)\}_{j=1}^{|J|}$ denote the sequence of assignment decisions produced by the current policy. After all passengers have received group assignments, the boarding simulation is run once to obtain the terminal reward $r^{|J|}$ defined in Eq.~\eqref{eq:two_obj}. Since intermediate rewards are zero, every decision in the episode shares the same return $G_b = r_b^{|J|}$.\par

Let $\theta_{\mathrm{old}}$ denote the actor parameters that generated the collected trajectories. For each decision $(s_b^j,a_b^j)$ in episode $b$, PPO uses the probability ratio:
\begin{equation}
    \rho_b^j(\theta)=
    \frac{\pi_\theta(a_b^j|s_b^j)}
    {\pi_{\theta_{\mathrm{old}}}(a_b^j|s_b^j)}.
\end{equation}
The advantage estimate is computed from the return and the critic baseline:
\begin{equation}
    A_b^j=G_b - V_\omega(s_b^j).
\end{equation}
The actor is trained with the clipped PPO surrogate objective:
\begin{equation}
    L^{\mathrm{clip}}(\theta)=
    \frac{1}{|\mathcal{D}|}\sum_{(b,j)\in\mathcal{D}}
    \min\left(
    \rho_b^j(\theta)A_b^j,\,
    \mathrm{clip}\left(\rho_b^j(\theta),1-\epsilon,1+\epsilon\right)A_b^j
    \right),
\end{equation}
where $\mathcal{D}$ is a mini-batch of assignment decisions and $\epsilon$ is the clipping parameter. The clipped ratio limits the size of each policy update, which is important in this problem because each trajectory contains many assignment decisions but only one terminal simulation outcome. An entropy bonus $\eta\,\mathcal{H}(\pi_\theta)$ is added to the actor objective to encourage exploration over boarding groups, where $\eta$ is the entropy coefficient.\par

The critic is trained to fit the terminal return:
\begin{equation}
    L^{V}(\omega)=
    \frac{1}{|\mathcal{D}|}\sum_{(b,j)\in\mathcal{D}}
    \left(G_b - V_\omega(s_b^j)\right)^2.
\end{equation}
In each training round, the collected trajectories are reused for multiple PPO epochs with shuffled mini-batches. The formulas above use raw episode returns for clarity; in the implementation, returns are normalized across each rollout batch before computing advantages and critic targets, and gradient clipping is applied when updating both networks.\par

Algorithm~\ref{alg:training} summarizes the PPO training procedure used to optimize the dynamic boarding policy.

\begin{algorithm}[H]
\caption{PPO Training for Dynamic Boarding Policy}
\label{alg:training}
\SetAlgoLined
Initialize actor parameters $\theta$ and critic parameters $\omega$\;
Set PPO clipping parameter $\epsilon$, entropy coefficient $\eta$, and number of PPO epochs $K_{\mathrm{PPO}}$\;
\While{not converged}{
    Collect an empty rollout buffer $\mathcal{B}$\;
	\For{$b=1$ \KwTo $B$}{
		Initialize an empty trajectory $\tau_b$\;
		\For{$j = 1$ to $|J|$}{
			Observe state $s_b^j = (x_b^j, y_b^j, z_b^j)$\;
            Sample action $a_b^j \sim \pi_\theta(\cdot | s_b^j)$\;
            Store $(s_b^j,a_b^j,\log \pi_\theta(a_b^j|s_b^j))$ in $\tau_b$\;
            Execute $a_b^j$ and observe the next state $s_b^{j+1}$\;
		}
		Simulate the boarding process and compute the normalized terminal reward $r_b^{|J|}$ via the reward equation above\;
		Set the episode return $G_b=r_b^{|J|}$ and attach $G_b$ to all decisions in $\tau_b$\;
        Add $\tau_b$ to $\mathcal{B}$\;
	}
	Normalize returns in $\mathcal{B}$: $\widehat{G}_b \leftarrow (G_b - \mu_G)/\sigma_G$, where $\mu_G$ and $\sigma_G$ are the mean and standard deviation of $\{G_b\}$ in $\mathcal{B}$\;
    Compute old log-probabilities, critic values, and advantages $A_b^j=\widehat{G}_b-V_\omega(s_b^j)$ for all decisions in $\mathcal{B}$\;
    \For{$k=1$ \KwTo $K_{\mathrm{PPO}}$}{
        Shuffle $\mathcal{B}$ and split it into mini-batches $\mathcal{D}$\;
        \For{each mini-batch $\mathcal{D}$}{
            Compute probability ratios $\rho_b^j(\theta)$ for all $(b,j)\in\mathcal{D}$\;
            Update $\theta$ by maximizing $L^{\mathrm{clip}}(\theta)+\eta\mathcal{H}(\pi_\theta)$\;
            Update $\omega$ by minimizing $L^V(\omega)$\;
        }
    }
}
\end{algorithm}

\section{Numerical Experiments}\label{sec:rl_evaluation}

This section evaluates the proposed dynamic RL policy against the static policies. Section~\ref{sec:rl_base_scenario} specifies the evaluation environment. Section~\ref{sec:static_policy_evaluation} benchmarks the three static policies from Section~\ref{sec:static_boarding_policies} across all six airplane layouts and identifies the strongest static competitor. Section~\ref{sec:rl_pareto} compares the RL policy against that strongest competitor and two luggage-ordering benchmarks (slow-first and fast-first) through a Pareto analysis of total boarding time and average individual boarding time. Section~\ref{sec:rl_robustness} tests whether the learned policy generalizes beyond the training setting. Section~\ref{sec:rl_layout_generalization} confirms that the RL policy remains the leading policy across all six cabin layouts.

\subsection{Experimental Setup}\label{sec:rl_base_scenario}

Travel-companion groups follow $(q_1,q_2,q_3)=(0.55,0.38,0.07)$, calibrated from the empirical boarding data in \citet{steiner2009speeding}. For carry-on luggage, we use a three-bin parameter set from the same source: $k\in\{0,1,2\}$ with probabilities $p_0=0.45$, $p_1=0.40$, and $p_2=0.15$; mean handling times $\mu_0=0$, $\mu_1=12.1$, and $\mu_2=25.3$~sec; and standard deviations $\sigma_0=0$, $\sigma_1=12.4$, and $\sigma_2=15.4$~sec.\footnote{The original three-piece luggage category is merged into the two-piece category.} The seat-interference parameter is $\xi=3.6$~sec. We report total boarding time and average individual boarding time as the two performance metrics, each as the mean over 1000 simulation replications. All policies are evaluated under common random seeds to ensure a fair comparison across policies. Regarding cabin layouts, the static policy benchmarks in Section~\ref{sec:static_policy_evaluation} cover all six cabin layouts in Section~\ref{sec:airplane_layout}. The RL experiments in Sections~\ref{sec:rl_pareto} and~\ref{sec:rl_robustness} use the $(3+3)\times 32$ single-aisle layout at full passenger load; Section~\ref{sec:rl_layout_generalization} then extends the RL evaluation to all six layouts.

\subsection{Static Policy Benchmarks}\label{sec:static_policy_evaluation}

Section~\ref{sec:experiment_two_bf} first studies the two-group back-to-front policy on two representative layouts, and Section~\ref{sec:comparison_four_static_policies} then compares all three static policies across all six layouts. All scenarios assume full passenger load, consistent with previous studies \citep{van2002reducing,steffen2008optimal,erland2024let}.

\subsubsection{Benchmarking the Two-Group Back-to-Front Policy}\label{sec:experiment_two_bf}

\begin{figure}[!b]
    \centering
    \includegraphics[width=0.92\textwidth]{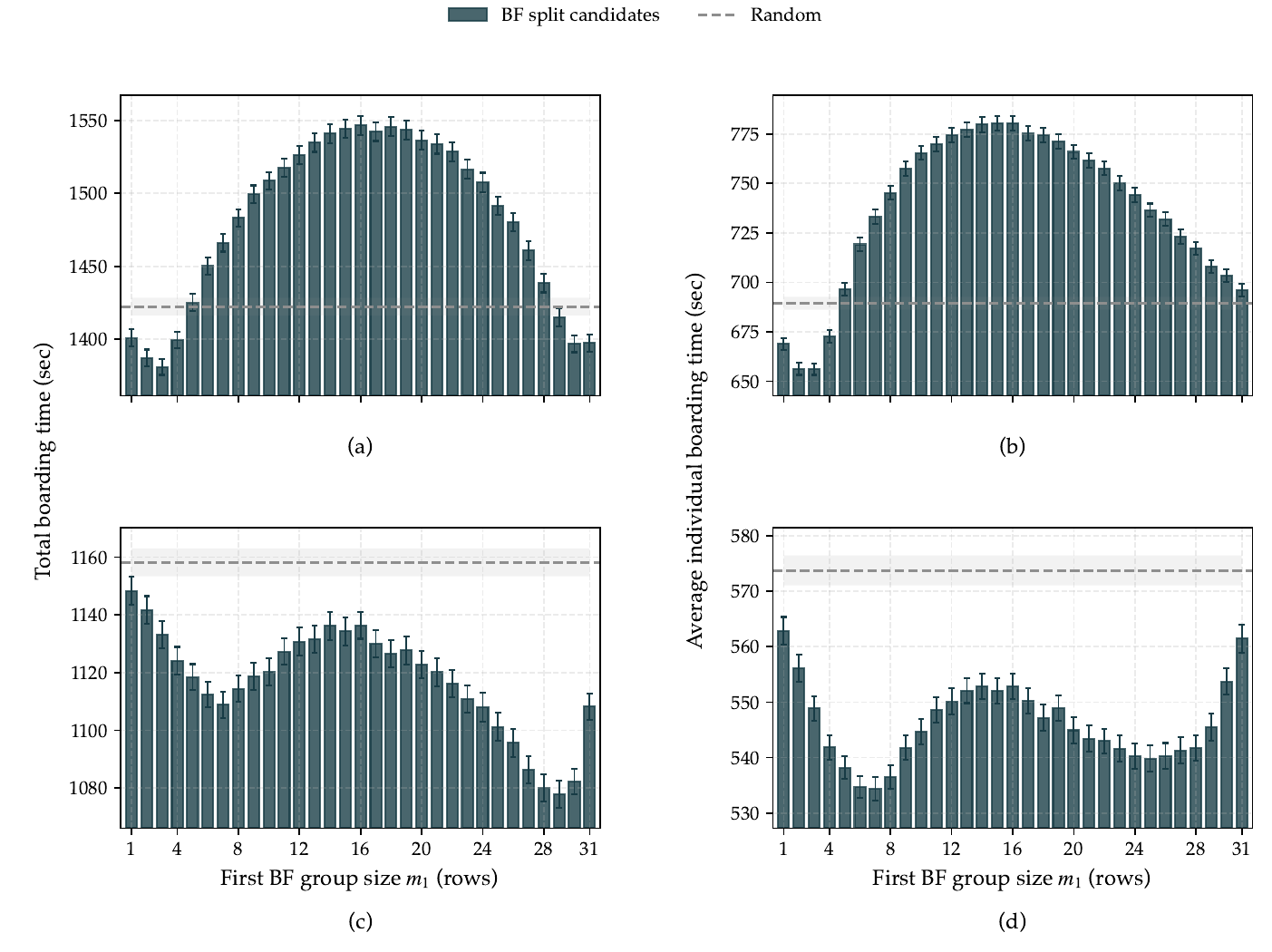}
    \caption{Two-group back-to-front row splits compared with random boarding: (a) total boarding time for the $(3+3)\times32$ layout; (b) average individual boarding time for the $(3+3)\times32$ layout; (c) total boarding time for the $(2+4+2)\times32$ layout; (d) average individual boarding time for the $(2+4+2)\times32$ layout.}
    \label{fig:section621_bf2_sensitivity}
\end{figure}

On the $(3+3)\times 32$ (single-aisle) and $(2+4+2)\times 32$ (double-aisle) layouts, we enumerate all possible row splits $[m_1, 32-m_1]$ and report total boarding time and average individual boarding time in Fig.~\ref{fig:section621_bf2_sensitivity}. In each panel, the bars give the mean boarding time over the 1000 replications for each split candidate, and the dashed line marks the random boarding baseline. Error bars and the shaded band around the baseline show 95\% confidence intervals for the mean.\par

First, we look at the total boarding time. Note that in both layouts, equal or near-equal splits produce particularly high total boarding time (Fig.~\ref{fig:section621_bf2_sensitivity}a and c). For the single-aisle case, these splits even exceed the random baseline, consistent with several previous studies that tested only equal splits \citep{ferrari2005improving,bachmat2009analysis,milne2014new}. The poor performance can be understood as follows. In the first stage, about half of all passengers head for the rear half of the cabin, and all luggage stowing and seat taking is confined to that half while the front half of the aisle carries only passengers walking through. In the second stage, the majority of the first group has already been seated and the second group occupies the front half, while the rear half of the aisle is now empty. Each stage therefore uses only about half of the cabin for parallel luggage stowing and seat taking. Random boarding avoids this problem by spreading passengers across the full cabin throughout the process.\par

However, highly unequal splits reverse the picture. The smaller group, whether it is the first or second one, has so few passengers and occupies so few rows that its members rarely interfere with the large group's boarding. This facilitates parallel boarding of the two groups. The total boarding time is therefore dominated by the large group, whose passengers are randomly ordered and spread across nearly the full cabin. The process is effectively a random boarding over a slightly smaller cabin with fewer rows and fewer passengers, which naturally takes less time. This finding is consistent with \citet{bachmat2013optimal}, who proved that two-group back-to-front outperforms random boarding only when the group sizes are sufficiently unequal.\par

In the simulated double-aisle layout, back-to-front outperforms random boarding across all row splits, including equal ones. The two aisles split the passenger flow, so each aisle serves fewer seats per row than the single aisle does. This lower congestion may let rear-bound passengers reach their rows faster and may shorten the unproductive occupation of the front aisle. Moreover, when one aisle is blocked by a passenger stowing luggage, the entry queue may continue feeding passengers into the other aisle, so the entry throughput is not completely interrupted.\par

For total boarding time, the double-aisle layout favors a large first group with a small second group (the large-first split), unlike the single-aisle layout, where the two extreme splits perform similarly and neither holds a clear advantage. To see why, it helps to view the total boarding time as the sum of two parts. The first part, $E_{\mathrm{total}}$, is the time needed for all passengers to enter the cabin; during this phase passengers are still moving along the aisle, so it is governed by how quickly the aisle clears, because a passenger can enter or move forward only when the space ahead is free. The second part, $D_{\mathrm{last}}$, is the extra time that the group boarding last (i.e., the front group) needs, after entering, to reach its seats and sit down; this is a closing phase that no later boarding can overlap with. The two parts respond to the row split in opposite directions. The total boarding time always includes the full seating time of the last group, so a small last group makes $D_{\mathrm{last}}$ short: the large-first split (the large group at the back, boarding first) places the small group last and keeps this closing phase short, whereas the small-first split places the large group last and makes it long. At the same time, $E_{\mathrm{total}}$ depends on how long passengers stay in the aisle before they sit. Under the large-first split the large group occupies the back rows, so most passengers walk a long way along a crowded aisle and leave it slowly, making $E_{\mathrm{total}}$ large; under the small-first split the large group occupies the front rows, so most passengers walk only a short way and leave the aisle quickly, making $E_{\mathrm{total}}$ small. The large-first split therefore shortens the closing phase but lengthens the entry phase, and the net result depends on how costly it is to send the large group to the back. In the single-aisle layout this cost is high, because the aisle is crowded and the walking distance is long; the longer entry phase of the large-first split then roughly cancels its shorter closing phase, and the two extreme splits give similar total boarding times, with no clear advantage. In the simulated double-aisle layout, the two aisles ease congestion, so the entry phase may grow only slightly under the large-first split; the shorter closing phase can then dominate, and the large-first split achieves the lower total boarding time.\par

For average individual boarding time, both layouts favor the small-first split, and the comparison is simpler than for total boarding time, because here the relevant effects reinforce each other instead of competing. The average is governed by the large group, since it contains most of the passengers. Two things determine when a passenger finally sits: the time spent waiting to enter the cabin, and the time spent reaching the assigned row and sitting down after entering. Placing the large group at the front shortens both. Its members walk only a short distance and sit soon after entering, which lowers their own boarding times; and because they leave the aisle quickly, the cabin fills faster, so passengers in general also enter and sit sooner. Both effects therefore favor placing the large group at the front, that is, the small-first split. There is no opposing effect here, because the average reflects the experience of a typical passenger rather than that of the last passenger to sit; the closing phase that matters for total boarding time does not affect the average in the same way. Consequently, the small-first split gives the lower average individual boarding time in both layouts. The size of this advantage follows the same congestion difference as before. In the single-aisle layout, sending the large group to the back is very costly, because the aisle is crowded and the walking distance is long, so avoiding it gives the small-first split a clear advantage. In the double-aisle layout the back rows are easier to reach, so the position of the large group matters less, and the advantage of the small-first split, although still present, is smaller.\par

The total-time and average-individual-time panels in Fig.~\ref{fig:section621_bf2_sensitivity} therefore convey a consistent message: two-group back-to-front is effective only when the row split is chosen carefully, and the split that minimizes total boarding time does not necessarily minimize average individual boarding time. (Appendix~\ref{apdx:additional_static} extends this analysis to three-group splits and shows that adding a third group can further reduce total boarding time when group sizes are chosen appropriately.) This observation motivates the two-objective evaluation in Section~\ref{sec:rl_pareto}.

\subsubsection{Comparison Across Cabin Layouts}\label{sec:comparison_four_static_policies}

\begin{figure}[!htb]
    \centering
    \includegraphics[width=0.92\textwidth]{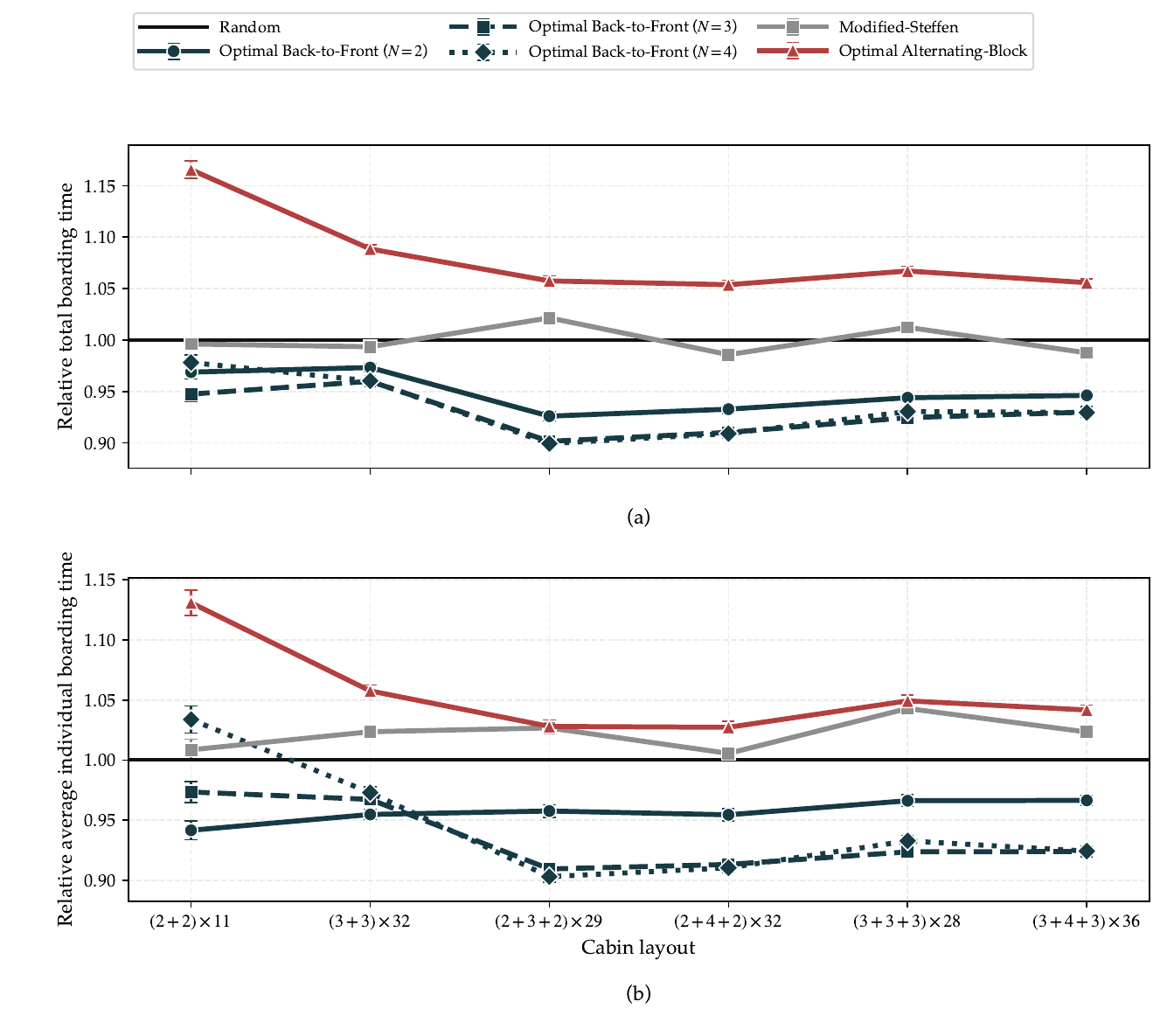}
    \caption{Companion-compatible static boarding policies compared across cabin layouts: (a) relative total boarding time; (b) relative average individual boarding time.}
    \label{fig:four_policy_comparison}
\end{figure}

We now compare the three static policies from Section~\ref{sec:static_boarding_policies} across the six airplane layouts in Section~\ref{sec:airplane_layout}. Fig.~\ref{fig:four_policy_comparison} reports each policy's total boarding time and average individual boarding time relative to random boarding: back-to-front with $N=2$, 3, and 4, modified-Steffen, and alternating-block. Values below one indicate improvement over random boarding. For back-to-front and alternating-block, we enumerate all feasible group size combinations and keep the one that minimizes total boarding time.

Across all six layouts and for every $N$, the optimal back-to-front policy beats random boarding on total boarding time; adding a third group lowers it further (Appendix~\ref{apdx:additional_static}), while including the fourth improves little. Back-to-front also beats random on average individual boarding time everywhere, except that the four-group policy is marginally worse than random on the small $(2+2)\times11$ cabin. In the simulated layouts, the improvement is larger in double-aisle cabins: four-group back-to-front cuts total boarding time by 7.0--10.1\% and average individual boarding time by 6.7--9.7\% relative to random boarding. Note that earlier works found no benefit from additional groups and did not model double-aisle cabins (e.g., \citealp{bachmat2013optimal}).\par

The modified-Steffen policy offers only marginal total-time improvement. In single-aisle layouts, it reduces total boarding time by less than 2\%, compared to the 6.2\% reported by \citet{steffen2008optimal}. The difference reflects our more detailed simulation that additionally accounts for seat interference, which also blocks the aisle and shrinks the portion of aisle blockage that spatial separation can reduce. In the $(2+3+2)$ and $(3+3+3)$ double-aisle layouts, modified-Steffen even increases total boarding time relative to random boarding. On average individual boarding time, it is worse than random in every layout.\par

The alternating-block policy consistently underperforms random boarding across all layouts on both metrics, even after selecting the best group sizes. The two mechanisms identified in Section~\ref{sec:static_boarding_policies}, later groups passing through unsettled blocks and large leading groups spilling forward, introduce additional aisle interference. This cost outweighs the spatial separation that the alternating design attempts to create between simultaneously boarding groups.

Among the static policies tested, the optimal back-to-front policy is the only one that consistently outperforms random boarding across all layouts and both metrics. Appendix~\ref{apdx:additional_static} further shows that the relative ordering of these policies remains stable as the number of rows varies, and that the back-to-front advantage grows with higher row counts.

\subsection{Evaluations of Dynamic RL Policy}\label{sec:rl_pareto}
\begin{figure}[!t]
    \centering
    \includegraphics[width=0.98\textwidth]{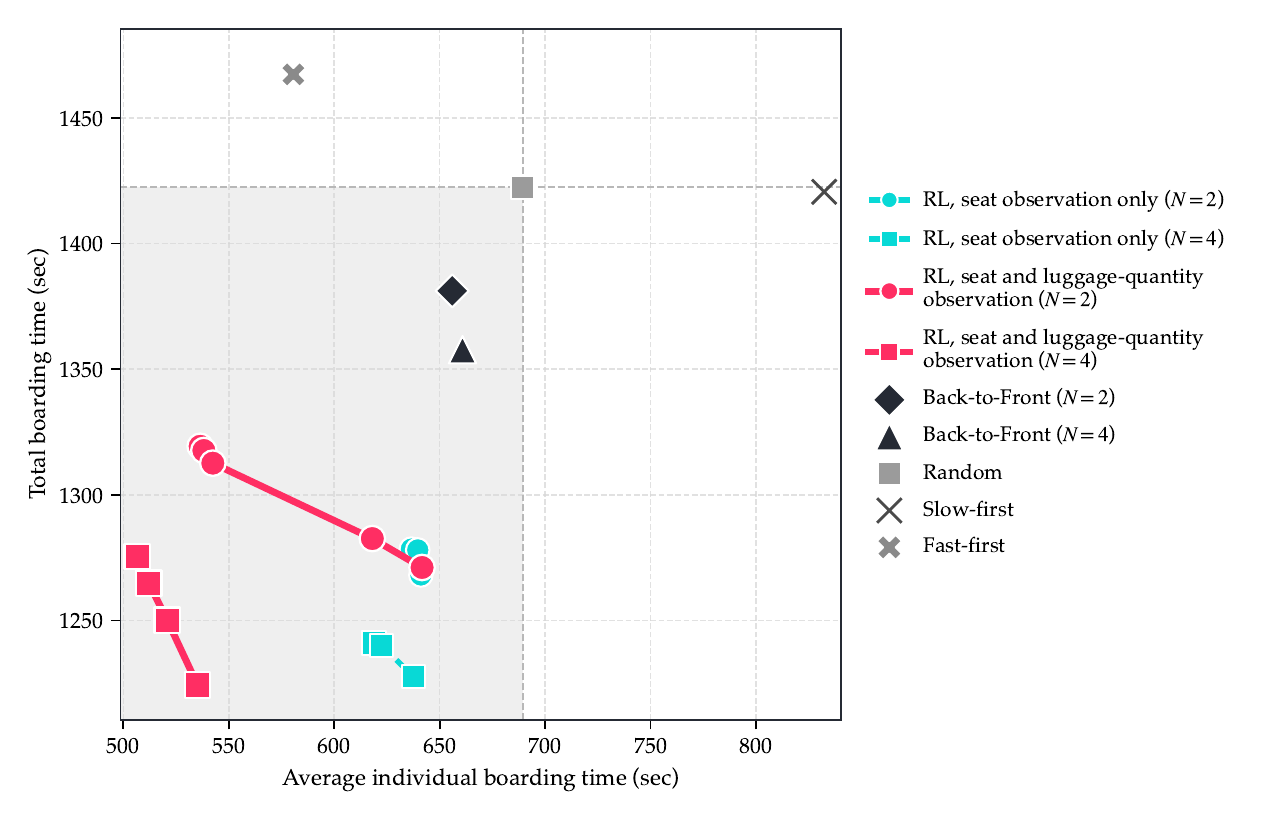}
    \caption{Pareto tradeoff between total boarding time and average individual boarding time on the $(3+3)\times32$ layout.}
    \label{fig:section72_unified_pareto}
\end{figure}

\citet{bachmat2023air} showed that under slow-first and fast-first luggage orderings, total boarding time and average individual boarding time can move in opposite directions. We therefore compare the RL policy against the optimal back-to-front policy from Section~\ref{sec:static_policy_evaluation}, random boarding as the baseline, and two luggage-ordering benchmarks (slow-first and fast-first), on both metrics under the $(3+3)\times32$ layout with the settings in Section~\ref{sec:rl_base_scenario}. We exclude the modified-Steffen and alternating-block policies, both of which perform generally worse than the optimal back-to-front in Section~\ref{sec:static_policy_evaluation}. Column-based policies such as outside-in and reverse-pyramid are excluded by construction, since they split companions across groups. Fig.~\ref{fig:section72_unified_pareto} presents the Pareto tradeoff between total and average individual boarding time; the policy settings that require further specification are as follows:

\begin{itemize}[leftmargin=*, itemsep=0pt, parsep=0pt, topsep=0pt, partopsep=0pt]
	\item \textit{RL policy with only seat observation.} The state observation includes seat information but not luggage quantities. Both $N=2$ and $N=4$ variants are trained for 6000 episodes. The normalized reward in Eq.~\eqref{eq:two_obj} is used, and $\lambda$ is swept from 0 to 1 in increments of 0.1 for each variant. We refer to these non-dominated evaluation points as an approximate Pareto frontier. Detailed network architecture and PPO hyperparameters are reported in Appendix~\ref{apdx:rl_implementation}.
    \item \textit{RL policy with seat and luggage observation.} Same as above, except that the observation also includes declared luggage quantities. Full declaration is assumed here; Section~\ref{sec:rl_robustness} relaxes this assumption.
	\item \textit{Back-to-front.} For both $N=2$ and $N=4$, all feasible row splits are enumerated. In each case, a single split dominates all others on both metrics, so the back-to-front frontier reduces to one point per $N$.
    \item \textit{Slow-first and fast-first.} Following \citet{bachmat2023air} and \citet{erland2024let}, slow-first boards companion groups with more declared luggage first, since more luggage implies longer stow time; fast-first uses the reverse order.
\end{itemize}

Fig.~\ref{fig:section72_unified_pareto} shows that the RL policies are generally Pareto-superior to the benchmarks; the shaded region marks outcomes that Pareto-dominate random boarding. The RL frontiers occupy the lower-left region, and contain both the lowest total boarding time and the lowest average individual boarding time among all compared policies. Quantitatively, relative to the four-group optimal back-to-front benchmark in Fig.~\ref{fig:section72_unified_pareto}, the two ends of the four-group RL frontier cut total boarding time by 9.8\% and average individual boarding time by 22.8\%, respectively. Two factors drive this advantage. First, the RL policy uses seat column information and still keeps companions in the same group. Static policies that preserve companion groups can only group by row; those that do use column information necessarily separate companions. Second, the policy is trained over many stochastic episodes, so it implicitly accounts for the randomness in passenger arrival order and luggage quantities. The attainable performance region therefore extends well beyond what companion-compatible static rules can reach.\par

The slow-first result is consistent with the mechanism discussed by \citet{bachmat2023air}: prioritizing passengers with longer luggage-handling times can slightly reduce total boarding time, but it increases average individual boarding time because these passengers enter the cabin earlier and remain there longer. In our simulation, however, the total-time improvement from slow-first is small, and the policy does not outperform the optimal back-to-front benchmarks. This suggests that luggage ordering alone is less effective at reducing total boarding time than row-based grouping with optimized group sizes.\par

The role of luggage information is more visible on average individual boarding time than on total boarding time. In Fig.~\ref{fig:section72_unified_pareto}, slow-first, fast-first, and the two RL variants (with and without luggage observation) are separated more along the horizontal axis than the vertical axis. This suggests that average individual boarding time is sensitive to where passengers with different luggage loads appear in the boarding sequence. Adding luggage observation to the RL state therefore substantially improves the average-individual-time side of the Pareto frontier. For $N=4$, the best average individual boarding time drops from 618.8 to 506.6 seconds (18.1\% reduction), while the minimum total boarding time changes only from 1227.7 to 1224.4 seconds.\par

\subsection{Robustness and Generalization}\label{sec:rl_robustness}

We next test whether the learned policy remains effective when operating conditions deviate from the training setting, without retraining. We evaluate the RL policy with $N=2$, seat and luggage observation, trained at $\lambda=0.2$. This $\lambda$ value sits near the minimum-total-time end of the frontier in Fig.~\ref{fig:section72_unified_pareto} but already noticeably improves average individual boarding time. We compare it with random boarding and the optimal back-to-front policy with $N=2$ from Section~\ref{sec:static_policy_evaluation}. All three policies are fixed throughout; none is re-optimized for the changed conditions.\par

We vary four aspects of the operating conditions:
\begin{itemize}[leftmargin=*, itemsep=0pt, parsep=0pt, topsep=0pt, partopsep=0pt]
    \item \textit{Passenger load factor.} The load factor takes values $50\%$, $60\%$, $70\%$, $80\%$, $90\%$, and $100\%$.
    \item \textit{Travel-companion distribution.} We vary $q=(q_1,q_2,q_3)$ across five settings with an increasing share of companion travelers: $(0.75,0.20,0.05)$, $(0.65,0.30,0.05)$, $(0.55,0.38,0.07)$, $(0.40,0.50,0.10)$, and $(0.30,0.50,0.20)$.
	\item \textit{Luggage distribution.} We test five cases with increasing luggage load, labeled \textit{Very light} through \textit{Very heavy}: $p=(0.65,0.30,0.05)$, $(0.55,0.35,0.10)$, $(0.45,0.40,0.15)$, $(0.35,0.43,0.22)$, and $(0.25,0.45,0.30)$. The nonzero luggage-handling times are scaled by $0.80$, $0.90$, $1.00$, $1.10$, and $1.20$ accordingly. 
	\item \textit{Luggage declaration ratio.} The declaration ratio takes values $0\%$, $25\%$, $50\%$, $75\%$, and $100\%$.
\end{itemize}

\begin{figure}[!tbp]
    \centering
    \includegraphics[width=0.9\textwidth]{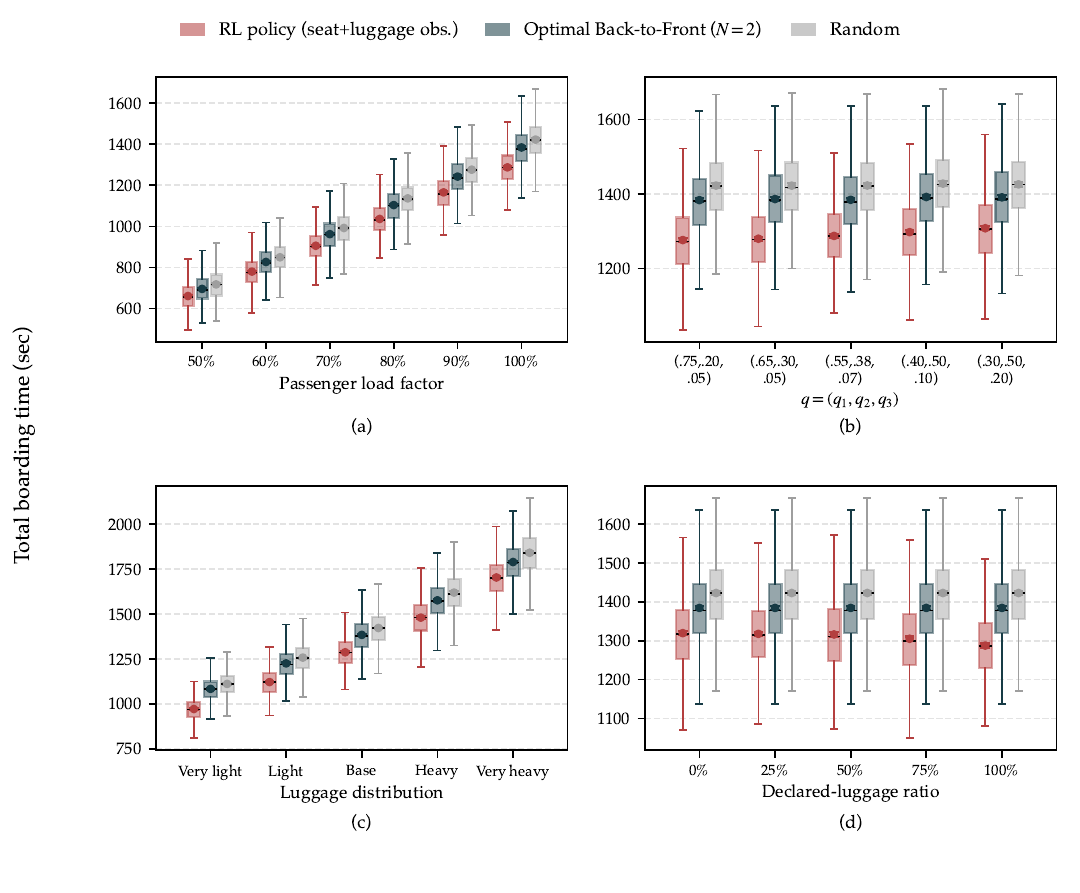}
    \caption{Robustness of total boarding time under changed operating conditions on the $(3+3)\times32$ layout: (a) passenger load factor; (b) travel-companion distribution; (c) luggage distribution; (d) luggage declaration ratio.}
    \label{fig:section73_robustness_total}
\end{figure}

\begin{figure}[!tbp]
    \centering
    \includegraphics[width=0.9\textwidth]{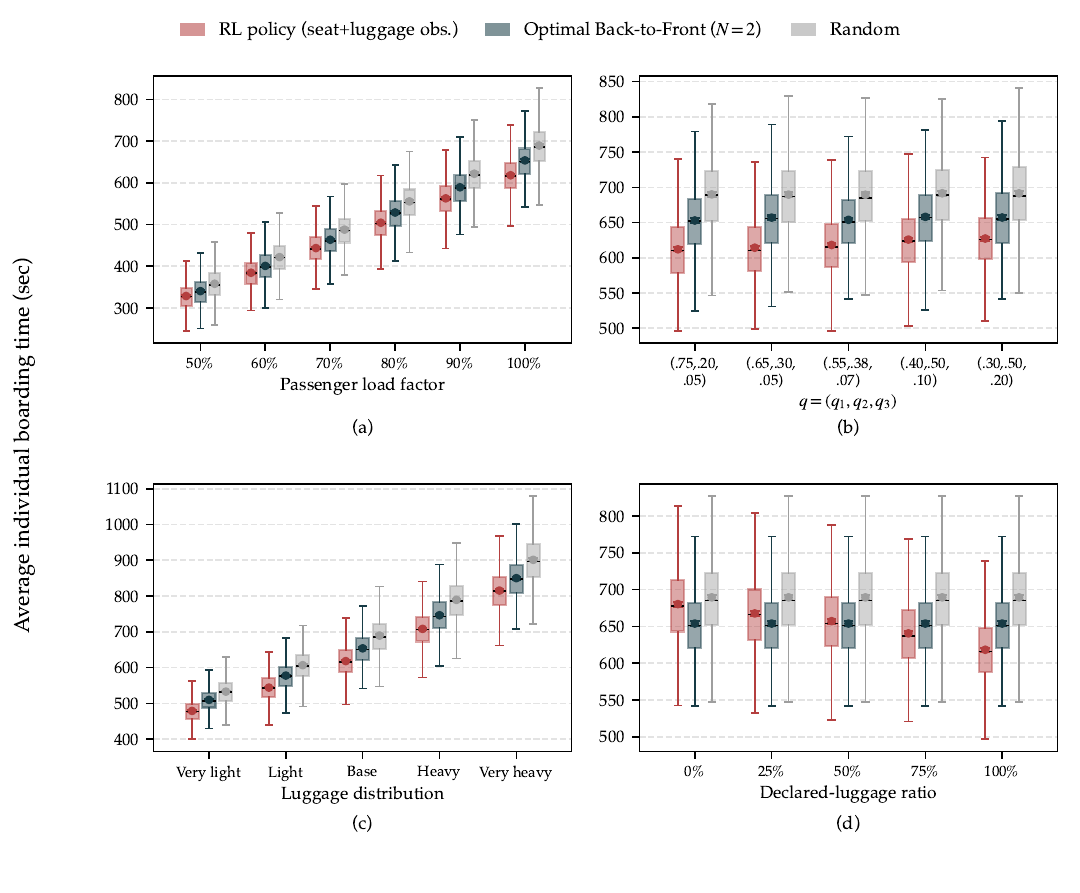}
    \caption{Robustness of average individual boarding time under changed operating conditions on the $(3+3)\times32$ layout: (a) passenger load factor; (b) travel-companion distribution; (c) luggage distribution; (d) luggage declaration ratio.}
    \label{fig:section73_robustness_avg}
\end{figure}

Figs.~\ref{fig:section73_robustness_total} and \ref{fig:section73_robustness_avg} report total boarding time and average individual boarding time, respectively, each computed from 1000 simulation runs per setting.\par

In Figs.~\ref{fig:section73_robustness_total}a and \ref{fig:section73_robustness_avg}a, the RL policy produces the lowest boarding time at every tested load factor. The three policies are close at low load factors, where sparse cabins limit the room for improvement. As the load factor increases, congestion grows and the RL policy's advantage widens, because denser cabins create more interference that the RL policy can mitigate.\par

For the travel-companion distribution test (Figs.~\ref{fig:section73_robustness_total}b and \ref{fig:section73_robustness_avg}b), larger companion groups force more passengers into the same boarding group and reduce the degrees of freedom in assignment. Despite this constraint, the RL policy consistently achieves the lowest total boarding time and average individual boarding time across all five settings.\par

The luggage distribution test (Figs.~\ref{fig:section73_robustness_total}c and \ref{fig:section73_robustness_avg}c) shows that heavier luggage loads raise both metrics for all policies, as expected from longer aisle-blocking times. The RL policy maintains the lowest total boarding time and average individual boarding time throughout.\par

The luggage declaration ratio test (Figs.~\ref{fig:section73_robustness_total}d and \ref{fig:section73_robustness_avg}d) is the only panel where the three policies respond differently. The back-to-front and random policies are unaffected by the declaration ratio, as neither uses luggage information. At 0\% declaration, the RL policy still beats both benchmarks on total boarding time using seat information alone, but its average individual boarding time is worse than the back-to-front benchmark.  As more passengers declare their luggage, the RL policy improves monotonically on both metrics, with larger gains on average individual boarding time than on total boarding time. This matches the Pareto analysis in Section~\ref{sec:rl_pareto}: seat-based assignment drives the total-time improvement, while luggage information is what mainly reduces average individual boarding time.\par

\subsection{RL Policy Evaluation Across Cabin Layouts}\label{sec:rl_layout_generalization}

The RL experiments in Sections~\ref{sec:rl_pareto} and~\ref{sec:rl_robustness} use the $(3+3)\times32$ layout. To test whether the same approach stays effective on other cabin layouts, we repeat the RL experiment on all six layouts in Section~\ref{sec:airplane_layout}: $(2+2)\times11$, $(3+3)\times32$, $(2+3+2)\times29$, $(2+4+2)\times32$, $(3+3+3)\times28$, and $(3+4+3)\times36$. For each layout we train a two-group RL policy with the same PPO hyperparameters, seat-plus-luggage observation, full luggage declaration, and objective weight $\lambda=0.2$, then compare it with random boarding and the optimal two-group back-to-front benchmark for the same layout using 1000 simulation runs.

\begin{figure}[htb!]
    \centering
    \includegraphics[width=0.9\textwidth]{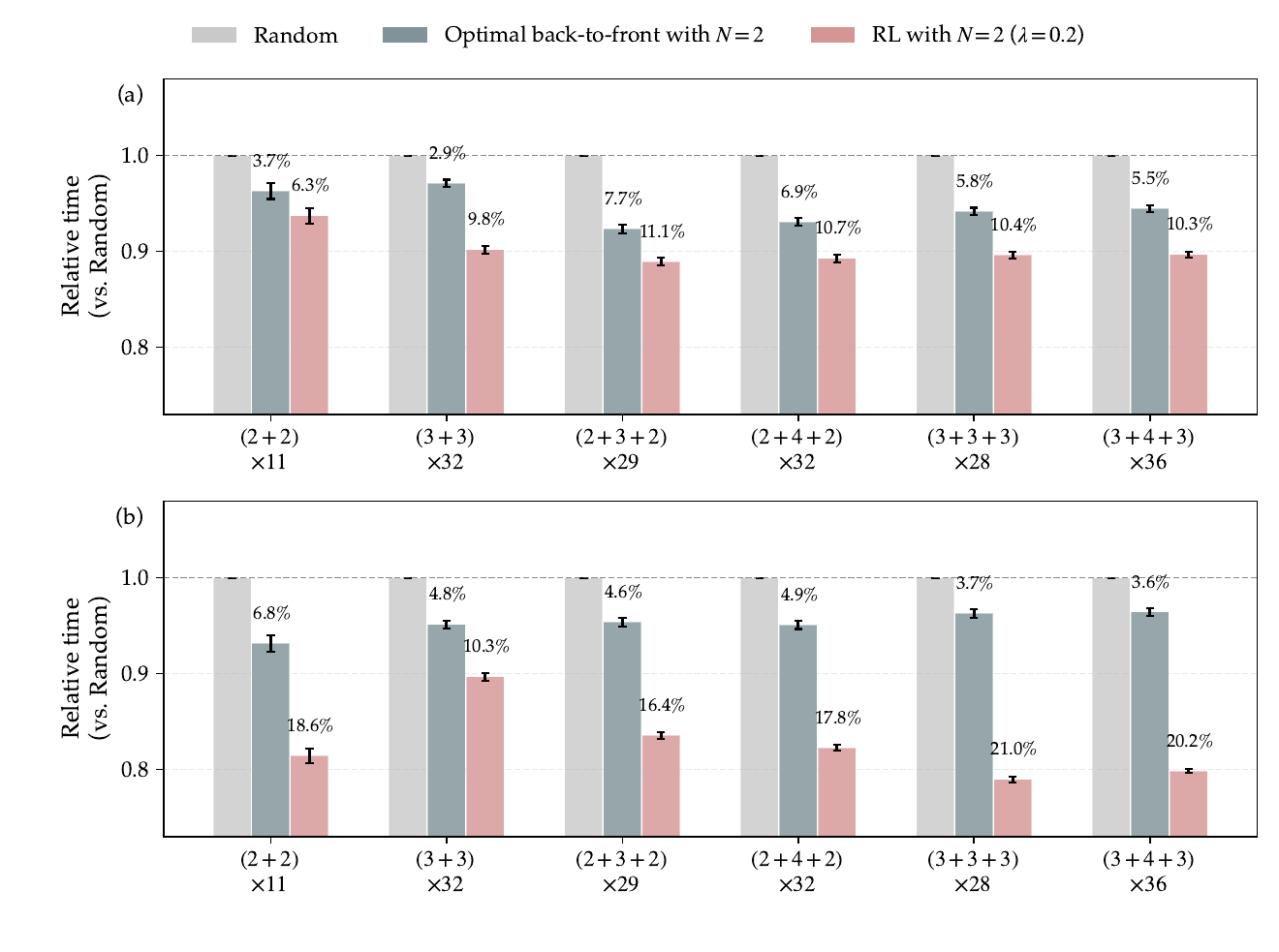}
    \caption{RL policy evaluated across cabin layouts against random boarding and the optimal two-group back-to-front: (a) relative total boarding time; (b) relative average individual boarding time.}
    \label{fig:layout_generalization}
\end{figure}

Fig.~\ref{fig:layout_generalization} reports each policy's total boarding time and average individual boarding time, in two panels, relative to random boarding within the same cabin layout. Across all six layouts, the three policies keep the same ranking as in Section~\ref{sec:rl_pareto}: the RL policy leads, the optimal two-group back-to-front follows, and random boarding trails. Specifically, compared with the optimal two-group back-to-front policy, the RL policy further reduces total boarding time by 2.7--7.1\% and average individual boarding time by 5.8--18.0\%, or 10--99 seconds and 24--130 seconds in absolute terms. Section~\ref{sec:comparison_four_static_policies} already showed that the optimal two-group back-to-front policy beats both modified-Steffen and alternating-block on both metrics in every layout, so the RL policy's lead over back-to-front implies that it outperforms all three static policies across all six layouts.\par

We extend the above analysis to $N=4$ and find that the ranking and general pattern between the RL and back-to-front policies hold across all six layouts; the results are omitted for brevity.

\FloatBarrier

\section{Conclusions}\label{sec:conclusions}

We consider a sequential boarding group assignment problem: passengers check in one at a time in no fixed order, and the airline must assign a group on arrival without seeing later passengers. In practice, airlines rely on fixed rules that map seat to a group. We formulate the assignment as an MDP for a commonplace setting where passengers are sorted into a few boarding groups and travel companions can board together, and solve it with a CNN-based RL policy trained by PPO. The policy exploits the seat column and optionally declared luggage quantities that companion-compatible static rules leave unused. To evaluate the policy, we develop a parsimonious simulation model calibrated from empirical boarding parameters that captures aisle interference, seat interference, and stochastic luggage handling. We then benchmark the learned policy against standard static policies across six representative single- and double-aisle economy-class layouts.\par

The experiments yield four main findings. First, among the three static policies tested, back-to-front with optimal group sizes is the strongest: it dominates modified-Steffen and alternating-block on both total and average individual boarding time across all layouts, and adding a third group further lowers both metrics while a fourth adds little. In the simulated layouts, the reduction relative to random boarding is larger on double-aisle cabins, reaching 7.0--10.1\% in total boarding time and 6.7--9.7\% in average individual boarding time. Second, the dynamic RL policy forms a Pareto frontier that lies below the optimal back-to-front, slow-first, and fast-first benchmarks across the total and average individual boarding time tradeoff on the representative $(3{+}3)\times32$ layout. The four-group RL frontier improves on the four-group optimal back-to-front benchmark by up to 9.8\% in total boarding time and up to 22.8\% in average individual boarding time at the two ends of the frontier. Across all six cabin layouts tested, the same RL approach also leads both metrics, so the advantage is not specific to the representative layout. Third, declared luggage quantities mainly improve average individual boarding time, while the seat column drives most of the reduction in total boarding time. Fourth, the learned policy is robust and generalizable: across changes in load factor, travel-companion distribution, luggage distribution, and luggage declaration ratio, a single trained policy largely outperforms optimal back-to-front without retraining.\par

Two recommendations follow for airline operations. Carriers that prefer a simple static rule should use back-to-front boarding with deliberately unequal group sizes: a two-group rule should use highly unequal groups, whereas a three-group rule should use small rear and front groups with a large middle group, especially for double-aisle aircraft in the tested simulation settings. Carriers that want the largest reductions in total boarding time or in average individual boarding time can embed the RL policy in the check-in process at counters or in mobile apps; the policy handles both solo travelers and companion groups, and asking passengers to declare luggage quantities at check-in yields a substantial additional gain in average individual boarding time.\par

That said, our study has three limitations that suggest future work. First, the simulation simplifies the real boarding process in several ways. Actual boarding involves late arrivals \citep{skorupski2015method}, state-dependent luggage stow times that vary with bin space and aisle-seat occupancy \citep{kierzkowski2017human}, and delays at boarding-pass checks \citep{kierzkowski2017human}. Second, although the simulation's operational parameters are calibrated from empirical data \citep{steiner2009speeding}, the model's outputs have not been validated against real boarding times. Future work could close this gap by aligning predicted total boarding times with field observations \citep{schultz2018field}. Third, like most RL applications, our approach learns from simulation, because the trial-and-error exploration it needs is too costly to run in live airport operations. Future work could transfer simulation-trained policies to real environments or learn directly from historical boarding data without live experimentation.

\section*{Acknowledgements}
This work was supported by the National Natural Science Foundation of China (Project Nos.~72571220 and 72201214), the Research Grants Council of Hong Kong (Project No.~15237624) and the Sichuan Science and Technology Program (Project No.~2025NSFSC1978). We also thank Ms.~Xiao Liang, a former undergraduate student at The Hong Kong Polytechnic University, whose Final Year Project work helped inspire this paper.

\clearpage
\begin{appendices}
\numberwithin{figure}{section}
\numberwithin{table}{section}
\numberwithin{algocf}{section}
\renewcommand{\theHfigure}{\thesection.\arabic{figure}}
\renewcommand{\theHtable}{\thesection.\arabic{table}}
\renewcommand{\theHalgocf}{\thesection.\arabic{algocf}}

\section{Table of Notations}\label{apdx:notation}
\begin{longtable}{@{}p{.15\textwidth}p{.8\textwidth}@{}}
\caption{List of notations}\label{table:notations}\\
\toprule
Notation & Description \\
\midrule
\endfirsthead
\caption[]{List of notations (continued)}\\
\toprule
Notation & Description \\
\midrule
\endhead
\bottomrule
\endlastfoot
$\mathcal{A}$ & Action space\\[0.07em]
$a^j$ & Action (group assignment) for passenger $j$\\[0.07em]
$A_b^j$ & Advantage estimate for decision $j$ in episode $b$\\[0.07em]
$\alpha_\theta$ & Learning rate for updating actor parameters\\[0.07em]
$\alpha_\omega$ & Learning rate for updating critic parameters\\[0.07em]
$B$ & Number of episodes collected in one PPO rollout batch\\[0.07em]
$C$ & Number of seats per row (column number)\\[0.07em]
$\mathcal{D}$ & Mini-batch of assignment decisions used in a PPO update\\[0.07em]
$D$ & Per-seat feature dimension ($N+5$)\\[0.07em]
$\Delta t$ & Time step size for discrete simulation\\[0.07em]
$e_x$ & Embedding of checked-in passenger states\\[0.07em]
$e_y$ & Embedding of current passenger state\\[0.07em]
$e_z$ & Embedding of global state\\[0.07em]
$\bar{e}$ & Concatenated embedding vector $[e_x; e_y; e_z]$\\[0.07em]
$\epsilon$ & PPO clipping parameter\\[0.07em]
$\eta$ & Entropy-bonus coefficient in PPO training\\[0.07em]
$F_i$ & Number of filters in $i$-th convolutional layer\\[0.07em]
$G$ & Episodic return\\[0.07em]
$G_b$ & Realized return for episode $b$\\[0.07em]
$\widehat{G}_b$ & Normalized return for episode $b$\\[0.07em]
$G_x$ & Dimension of checked-in state embedding\\[0.07em]
$G_y$ & Dimension of current passenger state embedding\\[0.07em]
$G_z$ & Dimension of global state embedding\\[0.07em]
$H$ & Dimension of final state representation\\[0.07em]
$h^j$ & Final state representation for passenger $j$ (subscripted $h^j_\theta$, $h^j_\omega$ for actor and critic)\\[0.07em]
$\mathcal{H}(\pi_\theta)$ & Entropy of the policy distribution\\[0.07em]
$J$ & Set of arriving passengers\\[0.07em]
$K_{\mathrm{PPO}}$ & Number of PPO epochs per rollout batch\\[0.07em]
$\boldsymbol{K}$ & Random variable representing number of luggage items per passenger\\[0.07em]
$\lambda$ & Weight in the reward function for the normalized $T_{\mathrm{total}}$ and $T_{\mathrm{avg}}$ tradeoff\\[0.07em]
$L$ & Total luggage-handling time for a passenger\\[0.07em]
$M$ & Maximum companion group size\\[0.07em]
$m_i$ & Number of rows assigned to group $i$ for back-to-front policy\\[0.07em]
$\mu_k$ & Mean total luggage-handling time if a passenger has $k$ items\\[0.07em]
$N$ & Number of boarding groups\\[0.07em]
$n(t)$ & Number of occupied seats in front of passenger's seat at time $t$\\[0.07em]
$n_i$ & Number of rows in group $i$ for alternating-block policy\\[0.07em]
$\boldsymbol{O}_k$ & Random variable of total luggage-handling time if a passenger has $k$ items \\[0.07em]
$o_k$ & Realized total luggage-handling time if a passenger has $k$ items\\[0.07em]
$\mathcal{P}$ & State transition dynamics\\[0.07em]
$p_k$ & Probability of a passenger having $k$ luggage items\\[0.07em]
$\phi(\cdot)$ & Activation function in the neural network\\[0.07em]
$\pi_\theta$ & Policy function parameterized by $\theta$\\[0.07em]
$q_k$ & Proportion of passengers traveling in companion groups of size $k$\\[0.07em]
$R$ & Total number of rows in the airplane\\[0.07em]
$\rho_b^j$ & PPO probability ratio for decision $j$ in episode $b$\\[0.07em]
$\mathcal{R}$ & Reward function\\[0.07em]
$r^j$ & Reward at step $j$\\[0.07em]
$\mathcal{S}$ & State space\\[0.07em]
$s^j$ & State observation for passenger $j$\\[0.07em]
$\sigma_k$ & Standard deviation of total luggage-handling time if a passenger has $k$ items\\[0.07em]
$T_{\mathrm{total}}$ & Total boarding time\\[0.07em]
$T_{\mathrm{avg}}$ & Average individual boarding time\\[0.07em]
$T_{\mathrm{total}}^{\mathrm{rand}}$ & Mean total boarding time under random boarding for the same layout and operating setting\\[0.07em]
$T_{\mathrm{avg}}^{\mathrm{rand}}$ & Mean average individual boarding time under random boarding for the same layout and operating setting\\[0.07em]
$V_\omega$ & Value function parameterized by $\omega$\\[0.07em]
$x^j$ & Checked-in state of previous passengers\\[0.07em]
$\Xi(\cdot)$ & State embedding function (subscripted $\Xi_\theta$, $\Xi_\omega$ for actor and critic)\\[0.07em]
$\xi$ & Basic time unit for passenger movement when taking seat\\[0.07em]
$y^j$ & Current passenger state\\[0.07em]
$z^j$ & Global state\\[0.07em]
\end{longtable}

\section{Additional Static-Policy Analyses}\label{apdx:additional_static}

\subsection{Benefits of Increasing Boarding Group Number for the Back-to-Front Policy}\label{apdx:bf_group_number}

\begin{figure}[!htb]
    \centering
    \includegraphics[width=0.78\textwidth]{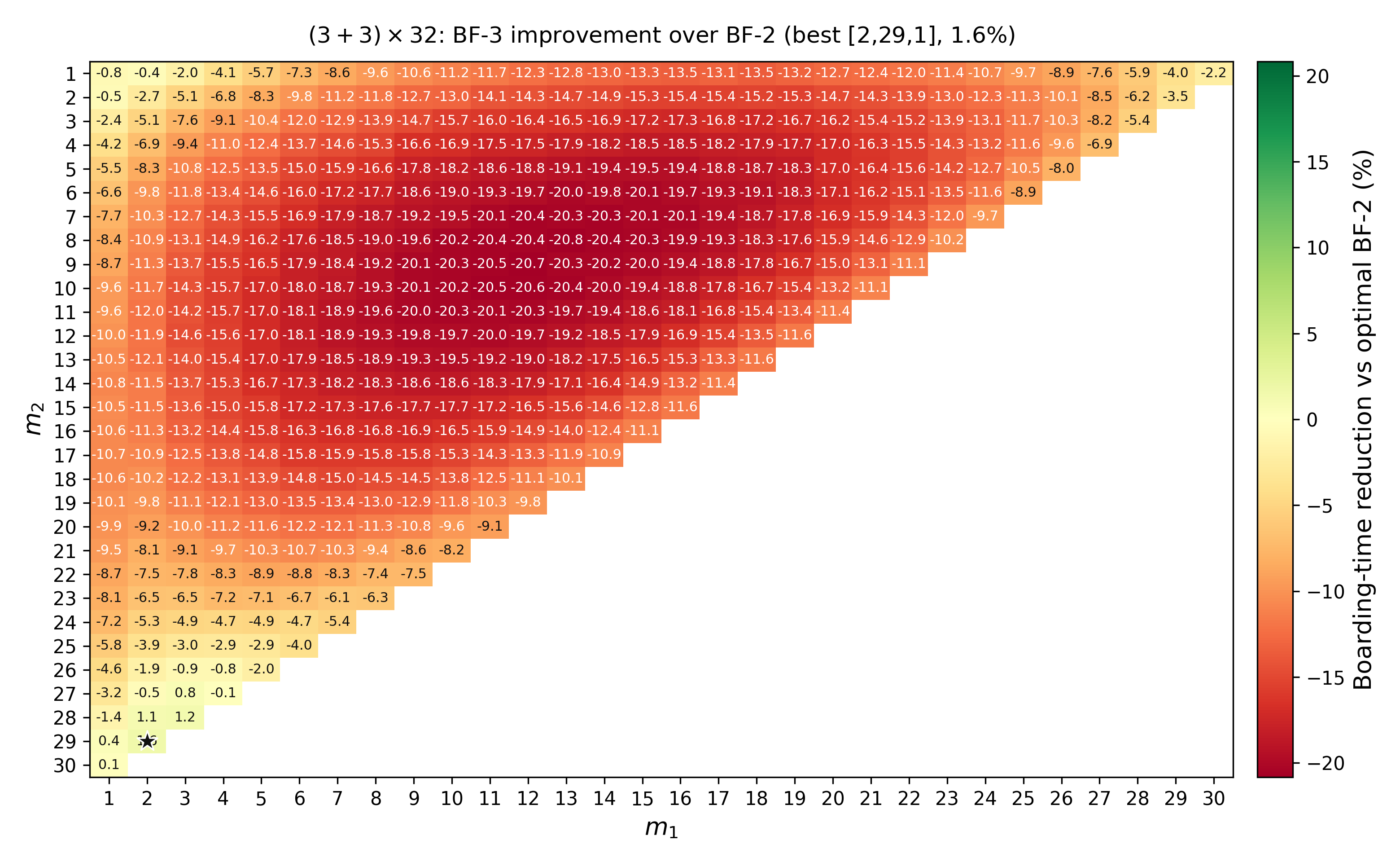}
    \caption{Three-group back-to-front performance relative to the optimal back-to-front policy with $N=2$ for the $(3+3)\times32$ single-aisle layout.}
    \label{fig:33_bf_enumerate_3_group}
\end{figure}

\begin{figure}[!htb]
    \centering
    \includegraphics[width=0.78\textwidth]{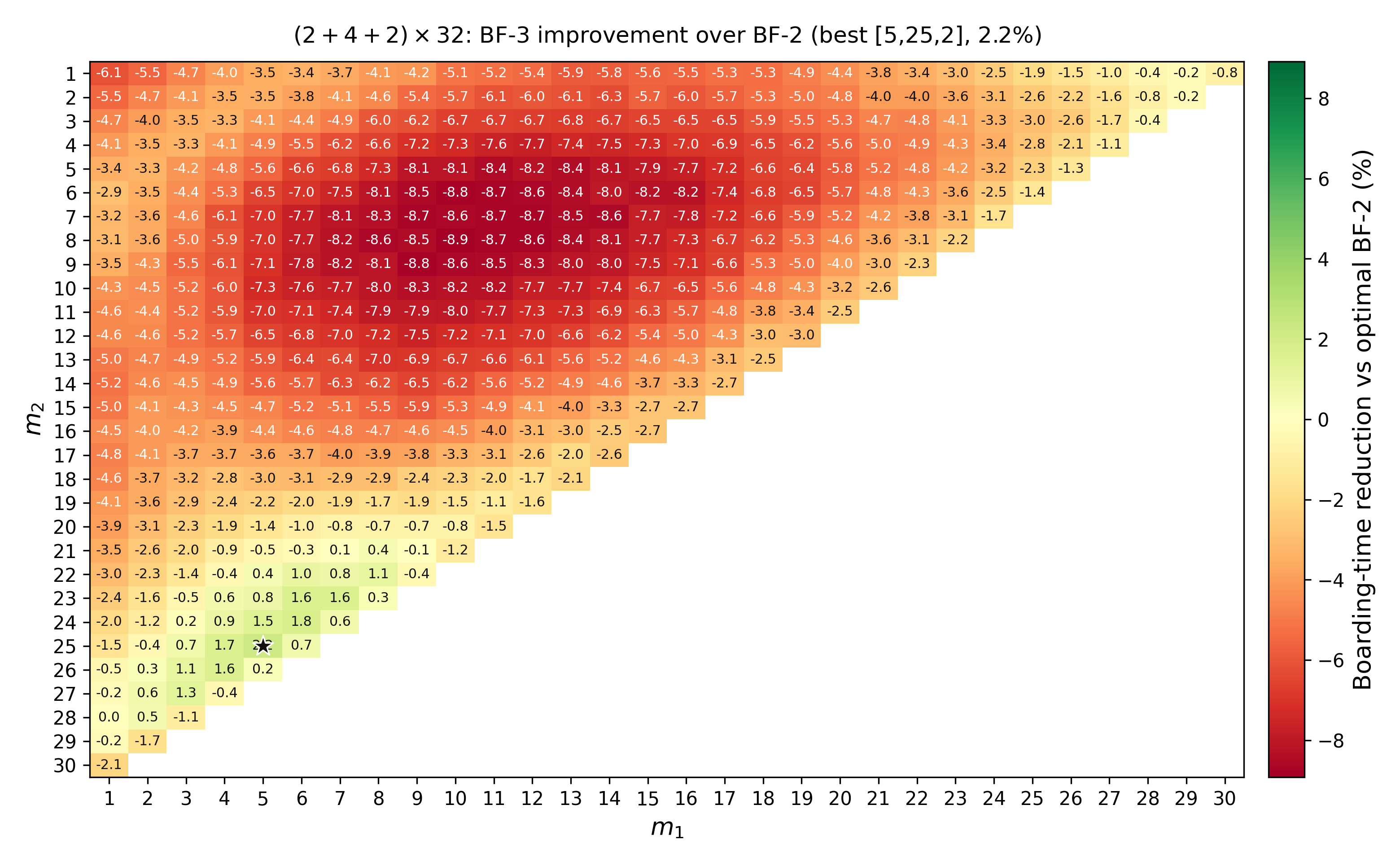}
    \caption{Three-group back-to-front performance relative to the optimal back-to-front policy with $N=2$ for the $(2+4+2)\times32$ double-aisle layout.}
    \label{fig:242_bf_enumerate_3_group}
\end{figure}

Figs.~\ref{fig:33_bf_enumerate_3_group} and \ref{fig:242_bf_enumerate_3_group} report total boarding time for three-group back-to-front policies on the $(3+3)\times32$ and $(2+4+2)\times32$ layouts, respectively, under the empirical luggage and travel-companion setting of the experimental setup in the main paper. We evaluate all possible combinations of group sizes $m_1$, $m_2$, and $m_3$. Due to the constraint $m_1+m_2+m_3=R$, each figure presents results in a two-dimensional heat map of $m_1$ and $m_2$ values. The value in each cell represents the percentage reduction in total boarding time relative to the \textit{optimal} two-group back-to-front policy under the same setting; greener cells indicate improvement and redder cells indicate deterioration.

Fig.~\ref{fig:33_bf_enumerate_3_group} reveals that the most effective arrangements feature a large middle group ($m_2$) with small first and last groups. The two-group back-to-front benchmark in the main paper showed that two-group back-to-front works best when group sizes are highly unequal, in either the small-first or large-first form; the three-group case effectively combines both, with a small group at each end and a large group in the middle.\par

For the double-aisle layout, Fig.~\ref{fig:242_bf_enumerate_3_group} shows a similar pattern, with the difference that the first group ($m_1$) can be moderately large while still achieving comparable reductions.\par

Adding a fourth group yields minimal improvements across all tested combinations, so we omit the detailed examination. Complete results comparing the $N=4$ case against other policies appear in the static-policy comparison in the main paper.

\subsection{Impacts of Row Numbers on Static-Policy Performance}\label{sec:sa_row_number}

\begin{figure}[!htb]
    \centering
    \includegraphics[width=0.82\textwidth]{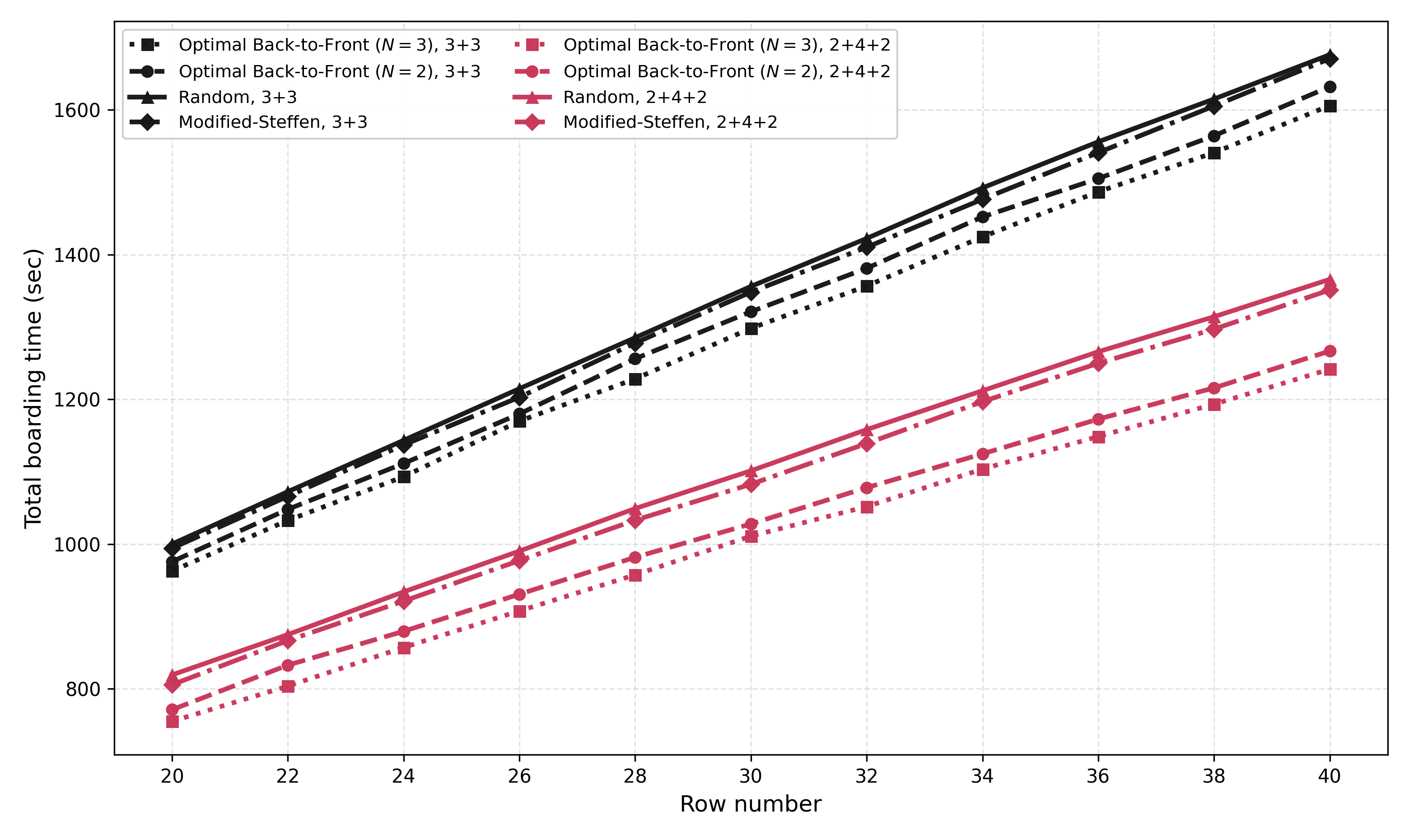}
    \caption{Total boarding time as a function of row number $R$ for the $(3+3)\times R$ and $(2+4+2)\times R$ layouts under the optimal back-to-front policies with $N=2$ and 3, the modified-Steffen policy, and the random policy.}
    \label{fig:row_impact}
\end{figure}

Fig.~\ref{fig:row_impact} shows how total boarding time varies with the row number $R$ under the $(3+3)\times R$ and $(2+4+2)\times R$ layouts for the optimal back-to-front policies with $N=2$ and 3, the modified-Steffen policy, and the random policy, under the empirical luggage and travel-companion setting of the experimental setup in the main paper. The back-to-front split is re-optimized for each row number. We exclude the alternating-block policy due to its consistently inferior performance.

All curves increase monotonically with the row number, because each additional row adds more passengers than the aisle can accommodate at once, so more passengers must queue outside the aisle and total boarding time grows. 

The ranking of the policies is consistent across row numbers, and the gaps between them remain roughly stable. Back-to-front therefore remains beneficial across the full range of row numbers, including the higher values typical of budget carriers.

\FloatBarrier

\clearpage
\section{RL Implementation Details}\label{apdx:rl_implementation}

Tables~\ref{tab:rl_network_architecture} and~\ref{tab:ppo_hyperparameters} summarize the neural-network architecture and PPO hyperparameters used in the Pareto evaluation. The actor and critic use the same embedding architecture but maintain separate parameters.

\begin{table}[H]
\centering
\caption{Neural-network hyperparameters used by the RL policy.}
\label{tab:rl_network_architecture}
\begin{tabularx}{\textwidth}{p{.34\textwidth}X}
\toprule
Hyperparameter & Value \\
\midrule
$F_1,F_2,F_3$ & $32,64,32$ \\
Convolution kernel & $3\times3$, padding 1 \\
$G_x$ & 64 \\
$G_y$ & 32 \\
$G_z$ & 16 \\
$H$ & 128 \\
$\phi(\cdot)$ & ReLU \\
\bottomrule
\end{tabularx}
\end{table}

\begin{table}[H]
\centering
\caption{PPO hyperparameters used for the RL policies in the Pareto evaluation.}
\label{tab:ppo_hyperparameters}
\begin{tabularx}{\textwidth}{p{.34\textwidth}X}
\toprule
Hyperparameter & Value \\
\midrule
Training episodes & 6000 \\
Optimizer & Adam \\
$\alpha_\theta,\alpha_\omega$ & $2.5\times 10^{-4}$ \\
$B$ & 5 \\
$|\mathcal{D}|$ & 64 \\
$K_{\mathrm{PPO}}$ & 4 \\
$\epsilon$ & 0.2 \\
$\eta$ & 0.01 \\
Gradient clipping norm & 0.5 \\
\bottomrule
\end{tabularx}
\end{table}

On the Mac Studio described in the simulation section of the main paper, a full 6000-episode PPO training run for the representative $(3+3)\times32$ setting completes in approximately 1 hour.

\end{appendices}

\clearpage
\begin{footnotesize}
\bibliographystyle{elsarticle-harv}
\bibliography{literature}
\end{footnotesize}

\end{document}